\documentclass[12pt]{iopart}

\usepackage{graphicx}
\begin{document}

\title[Transport and localization properties of excitations in one-dimensional lattices]{Transport and localization properties of excitations in one-dimensional lattices with
diagonal disordered mosaic modulations}

\author{Ba Phi Nguyen$^{1,2}$  and Kihong Kim$^{3,4}$\footnote{Author to whom any correspondence should be addressed}}

\address{$1$
Department of Basic Sciences, Mientrung University of Civil Engineering, Tuy Hoa 620000, Vietnam\\$2$ Research Institute for Basic Sciences, Ajou University, Suwon 16499, Korea\\$3$  Department of Physics, Ajou University, Suwon 16499, Korea\\$4$ School of Physics, Korea Institute for Advanced Study, Seoul 02455, Korea}
\ead{khkim@ajou.ac.kr}

\vspace{10pt}
\begin{indented}
\item[]July 2023
\end{indented}

\begin{abstract}
We present a numerical study of the transport and localization properties of excitations in one-dimensional lattices with diagonal disordered mosaic modulations. The model is characterized by the modulation period $\kappa$ and the disorder strength $W$. We calculate the disorder averages $\langle T\rangle$, $\langle \ln T\rangle$, and $\langle P\rangle$, where $T$ is the transmittance and $P$ is the participation ratio, as a function of
energy $E$ and system size $L$, for different values of $\kappa$ and $W$. For excitations at quasiresonance energies determined by $\kappa$, we find power-law scaling behaviors of the form $\langle T \rangle \propto L^{-\gamma_{a}}$,
$\langle \ln T \rangle \approx -\gamma_g \ln L$, and $\langle P \rangle \propto L^{\beta}$, as $L$ increases to a large value.
In the strong disorder limit, the exponents are seen to saturate at the values $\gamma_a \sim 0.5$, $\gamma_g \sim 1$, and $\beta\sim 0.3$,
regardless of the quasiresonance energy value. This behavior is in contrast to the exponential
localization behavior occurring at all other energies.
The appearance of sharp peaks in the participation ratio spectrum at quasiresonance energies provides additional evidence for the existence of an anomalous power-law localization phenomenon.
The corresponding eigenstates demonstrate multifractal behavior and exhibit unique node structures.
In addition, we investigate the time-dependent wave packet dynamics and calculate the mean square displacement $\langle m^2(t) \rangle$, spatial probability distribution, participation number, and return probability.
When the wave packet's initial momentum satisfies the quasiresonance condition, we
observe a subdiffusive spreading of the wave packet, characterized by $\langle m^2(t) \rangle\propto t^{\eta}$ where $\eta$ is always less than 1.
We also note the occurrence of partial localization at quasiresonance energies, as indicated by the saturation of the participation number and a nonzero value for the return probability at long times.
\end{abstract}

%

\submitto{\JPA}
%
%
%

\section{Introduction}
\label{sec1}
According to the scaling theory of Anderson localization in noninteracting systems, localization occurs in one- and two-dimensional systems with arbitrarily weak disorder, and in three-dimensional systems with strong enough disorder \cite{Abr,Lee0}. Despite extensive studies on Anderson localization \cite{And,Lif,She,Eve,Lag,Izra}, there are still several aspects that are not fully understood, and new types of localization phenomena continue to be discovered \cite{Fal,Ama1,Ama2,Saha,Raz,Basiri,Ngu4,Tzo,Wei,Wang,Ngu5,KK1,Del1,Del2,Del3,Noro,Noro2,Ley,Kim1,Kim2,Kim3,Kim4,Cros1,Cros2,Ber}. Recent examples include unconventional transport in disordered quantum wires \cite{Fal,Ama1,Ama2,Saha,Raz}, localization in disordered non-Hermitian systems \cite{Basiri,Ngu4,Tzo,Wei,Wang,Ngu5,KK1}, and the quantum boomerang effect in both Hermitian and non-Hermitian systems \cite{Del1,Del2,Del3,Noro,Noro2}.
Furthermore, the effect of interactions on Anderson localization and the related phenomenon of many-body localization has been an important and challenging research topic for many decades
that still raises many open questions \cite{Basko,Dias,Abanin}.

It is well-known that in one-dimensional (1D) noninteracting systems with uncorrelated disorder, long-range transport is absent due to Anderson localization \cite{And}.  This can be observed through an exponential decrease in transmittance with thickness  for extended excitations, and a saturation of mean-square displacement at long times for initially localized excitations. However, studies have shown that certain types of correlations in the disorder distribution can enable long-range transport. For example, the binary random dimer model exhibits a discrete set of resonance energies \cite{Dun,Wu,Dat,Izra1,Nae}, where the mean-square displacement grows as $t^{3/2}$ over time $t$, indicating superdiffusive transport behavior \cite{Dat}. Additionally, the transmittance across the system is identically equal to 1 in the scattering problem. This superdiffusive behavior has been observed in optical experiments \cite{Nae}. The binary random dimer model has been extended to the binary random $N$-mer model, and analytical expressions have been derived to identify the resonant energies that trigger delocalization \cite{Kos1, Kos2}.

Our recent numerical investigation explored the time-dependent reflection of wave packets incident on an effectively semi-infinite disordered mosaic lattice chain, where disordered on-site potentials are inserted into the lattice only at equally spaced sites \cite{Ngu0}. Through extensive numerical calculations, we discovered a discrete set of quasiresonance energies that deviate sharply from the ordinary Anderson localization behavior. We derived a simple analytical formula for these energies, which interestingly takes the same form as the resonance energy formula found in the binary random $N$-mer model \cite{Kos1}. However, it is important to note that this similarity is purely coincidental, as the underlying mechanisms are entirely distinct  \cite{Ngu0}. The binary random $N$-mer model exhibits superdiffusive wave packet spreading and completely extended states at resonance energies. Therefore, a natural question arises: how do these phenomena change at the quasiresonance energies present in the disordered mosaic lattice model? One of our primary goals is to address this question in the present study.
It is worth noting that 1D mosaic lattice models of various types have yielded many interesting results in recent years \cite{Wan1,Zen1, Zen,Liu, Gon, Dwi, Zen2}.

Recently, we have learned that the model known as the diluted Anderson model, which is essentially identical to the disordered mosaic lattice model discussed in this paper, was first studied by Hilke and later explored by various research groups \cite{Hilke,Moura}. Nonetheless, our work deviates significantly from previous studies and presents a multitude of fresh and interesting results.
In \cite{Hilke}, it was demonstrated that the underlying periodicity of the lattice gives rise to resonance energies. Although the resonance energies have been correctly derived, there has been a mischaracterization regarding the nature of the resonant states, incorrectly labeling them as entirely extended states.
In both this paper and our previous work \cite{Ngu0}, we provide unambiguous evidence through calculations of various physical quantities that these states are not entirely extended; rather, they exhibit power-law localized critical behavior.
de Moura {\it et al.} investigated the dynamics of a wave packet launched in a diluted Anderson lattice and observed
a subdiffusive spreading characterized by the mean square displacement behavior $\langle m^2(t)\rangle\propto t^{0.5}$ \cite{Moura}.
However, their study focused on an initially localized wave packet represented by a single site $\delta$ function.
As a result, it encompassed contributions from all eigenstates, including both the localized and quasiresonant states, thereby making it challenging to distinguish the energy dependencies of the wave packet dynamics.
Contrastingly, our present work employs a Gaussian wave packet with a narrow momentum distribution, enabling us to specifically investigate the distinct influence of the quasiresonant states on the dynamics of the wave packet.

In this paper, we aim to further expand our previous research
and investigate the nature of states at quasiresonance energies in more detail through calculations of several other physical quantities. Specifically, we study three different characteristics. First, we calculate the disorder averages of the transmittance and the logarithm of transmittance, denoted as $\langle T\rangle$ and $\langle\ln T\rangle$, in the scattering geometry and perform a finite-size scaling analysis of the results. Second, we calculate the averaged participation ratio $\langle P\rangle$ by solving the eigenvalue problem. For excitations at quasiresonance energies, we find power-law scaling behaviors of the form $\langle T \rangle \propto L^{-\gamma_{a}}$, $\langle \ln T \rangle \approx -\gamma_g \ln L$, and $\langle P \rangle \propto L^{\beta}$, as the system length $L$ increases to a large value. This behavior is in stark contrast to the exponential localization behavior displayed at all other energies.
The appearance of sharp peaks in the participation ratio spectrum at quasiresonance energies lends further support to the notion of an anomalous power-law localization effect. The corresponding eigenstates demonstrate multifractal behavior and exhibit unique node structures.
Thirdly, we investigate the time-dependent wave packet dynamics and calculate the mean square displacement $\langle m^2(t) \rangle$, spatial probability distribution, participation number, and return probability.
When the wave packet's initial momentum satisfies the quasiresonance condition, we consistently
observe a subdiffusive spreading of the wave packet, characterized by $\langle m^2(t) \rangle\propto t^{\eta}$ where $0<\eta<1$.
Furthermore, we note the occurrence of partial localization at quasiresonance energies, as indicated by the saturation of the participation number and a nonzero value for the return probability at long times.

The rest of this paper is organized as follows. In section \ref{sec2}, we introduce the 1D disordered mosaic lattice model. The results
of the numerical calculations are presented in the order of wave transmittance, participation ratio, and time-dependent wave packet dynamics in section \ref{sec3}. Finally, we conclude the paper in section \ref{sec41}.

\section{Model}
\label{sec2}
We consider a quantum particle that moves along a 1D lattice, which can be characterized by the time-dependent discrete Schr\"odinger equation
\begin{eqnarray}
i\hbar\frac{dC_{n}(t)}{dt}=V_{n}C_{n}(t)+J[C_{n-1}(t)+C_{n+1}(t)],
\label{equation1}
\end{eqnarray}
where $C_{n} (t)$ is the probability amplitude for finding the particle at the $n$-th site at time $t$, subject to the normalization condition $\sum_{n}\vert C_{n}(t)\vert^2=1$. $V_{n}$ is the on-site potential at the $n$-th site and $J$ is the coupling strength between adjacent sites. From now on, we measure all energy scales in the units of $J$ and set $J=\hbar=1$, which implies that the energy coincides with the frequency. The stationary solutions of equation~(\ref{equation1}) can be represented in the conventional form, $C_{n}(t)=\psi_{n}e^{-iEt}$, where $E$ is the energy of an eigenstate. Then we obtain
\begin{eqnarray}
E{\psi_{n}}=V_{n}\psi_{n}+ \psi_{n-1}+\psi_{n+1}.
\label{equation2}
\end{eqnarray}

\begin{figure}
\centering
\includegraphics[width=10cm]{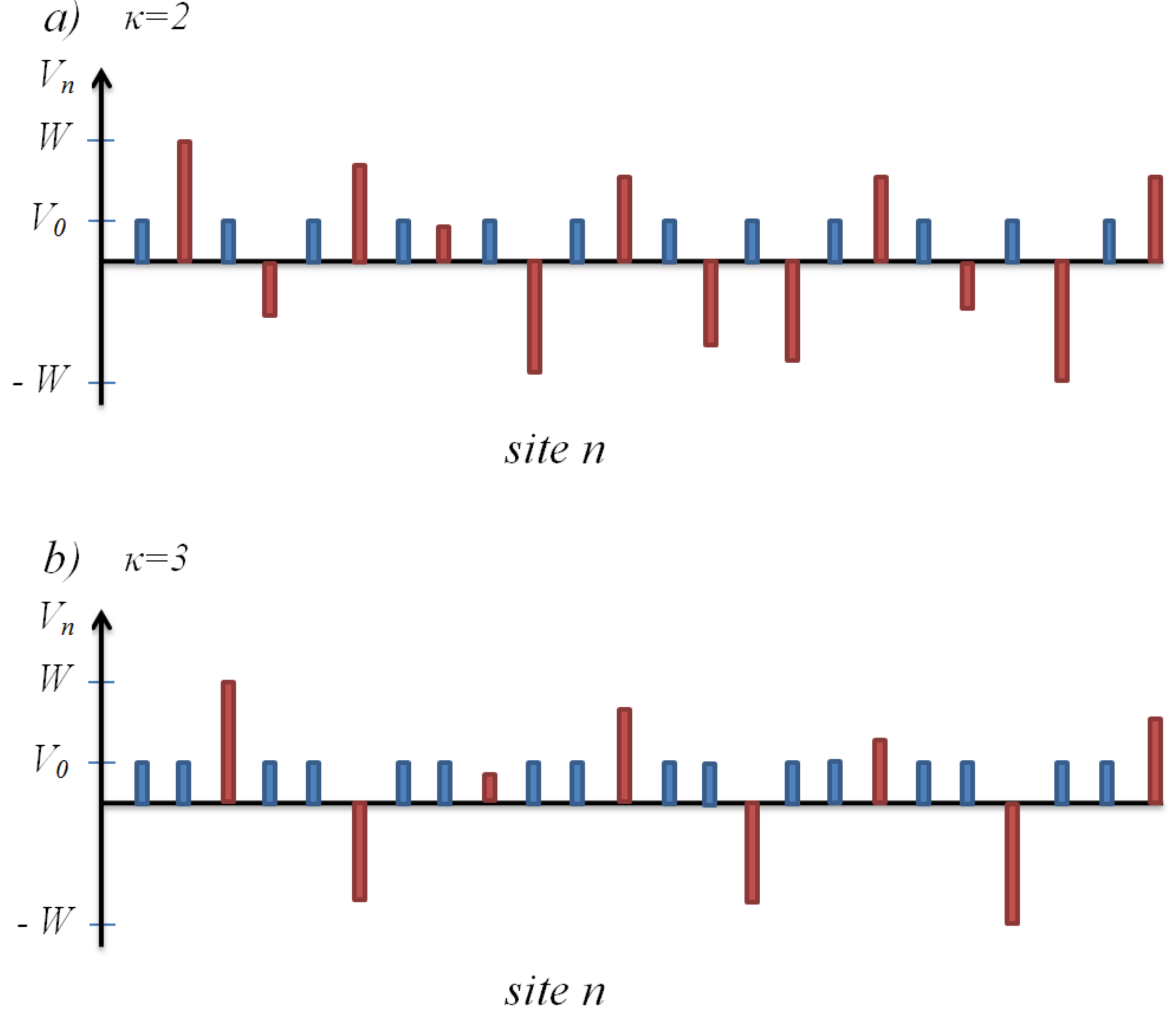}
\caption{Examples illustrating the typical spatial profile of the on-site potential in the 1D disordered mosaic lattice model for (a) $\kappa=2$ and (b) $\kappa=3$.}
\label{fig0}
\end{figure}

In the present study, we will investigate the transport and localization properties of a disordered mosaic lattice model
characterized by an on-site potential of the following form:
\begin{eqnarray}
V_{n}=\left\{\begin{array}{l l}
\varepsilon_{n}\in[-W,W], & \quad n=m\kappa\\
V_{0}, & \quad \mbox{otherwise}
\end{array}\right.,
\label{equation3}
\end{eqnarray}
where $\kappa$ is a positive integer referred to as the mosaic modulation period and $m$ is an integer ranging from $1$ to $N$.
This equation specifies that the on-site potential undergoes mosaic modulation with a periodicity of $\kappa$.
The total number of sites $L$ is equal to $\kappa N$. The on-site potential $\varepsilon_{n}$ at the $m\kappa$-th site is a uniformly distributed random variable over the interval $[-W, W]$, where $W$ represents the magnitude of disorder. At all the other sites, the on-site potential remains constant with a value of $V_{0}$.
The disordered mosaic lattice model has been the subject of recent investigation, with emphasis on numerical analysis of the time-dependent reflectance of wave packets incident upon a lattice with a large length $L$ \cite{Ngu0}.
The current study significantly expands on previous research by exploring additional aspects of the transport and localization properties of the same model, within the context of both stationary and non-stationary problems.
In figure~\ref{fig0}, we illustrate the typical spatial profile of the on-site potential in the 1D disordered mosaic lattice model for $\kappa=2$ and 3.

\section{Numerical results}
\label{sec3}

In this section, we provide comprehensive numerical results in the order of wave transmittance, participation ratio, and time-dependent wave packet dynamics.

\subsection{Wave transmittance}
\label{sec31}

We assume that a plane wave is incident from the right side of a 1D lattice chain of length $L$ and define
the amplitudes of the incident, reflected, and transmitted waves, $r_{0}$, $r_{1}$, and $t$, by
\begin{eqnarray}
\psi_{n}=\left\{\begin{array}{l l}
r_{0}e^{-iq(n-L)}+r_{1}e^{iq(n-L)}, & \quad \mbox{$n\geq L$}\\
te^{-iqn}, & \quad \mbox{$n\leq 0$}
\end{array}\right.,
\label{equation4}
\end{eqnarray}
where the wave number $q$ is related to $E$ by the free-space dispersion relation $E=2\cos q$. In the absence of dissipation, the conservation law $|r_{1}|^2+|t|^2=|r_{0}|^2$ is satisfied. We choose the overall constant phase for the wave functions such that $t$ corresponds to a positive real number.

The transmission coefficient is a crucial parameter for determining the nature of the states, and in experiments, it is associated with the dc conductivity of the system being studied.
In order to compute the transmittance and reflectance numerically,
we begin by selecting a positive real number for $t$ arbitrarily.
Then we use the relationships $\psi_{-1}=te^{iq}$, $\psi_{0}=t$, and $\psi_{1}=E\psi_{0}-\psi_{-1}$ and solve equation~(\ref{equation2}) iteratively until we obtain $\psi_{L}$ and $\psi_{L+1}$. Next, using the relationships $\psi_{L}=r_{0}+r_{1}$ and $\psi_{L+1}=r_{0}e^{-iq}+r_{1}e^{iq}$, we compute
\begin{eqnarray}
r_{0}=\frac{\psi_{L}e^{iq}-\psi_{L+1}}{e^{iq}-e^{-iq}},~r_{1}=\frac{-\psi_{L}e^{-iq}+\psi_{L+1}}{e^{iq}-e^{-iq}}.
\label{equation5}
\end{eqnarray}
Finally, the transmittance $T$ and the reflectance $R$ are obtained from
\begin{eqnarray}
&&{T}(E)=\left\vert\frac{t}{r_{0}}\right\vert^2=\vert t\vert^2\frac{4\sin^{2}q}{\left\vert\psi_{L}e^{iq}-\psi_{L+1}\right\vert^2},
\label{equation6}\\
&&{R}(E)=\left\vert\frac{r_{1}}{r_{0}}\right\vert^2=\left\vert\frac{\psi_{L}e^{-iq}-\psi_{L+1}}{\psi_{L}e^{iq}-\psi_{L+1}}\right\vert^2.
\label{equation7}
\end{eqnarray}

\begin{figure}
\centering
\includegraphics[width=12cm]{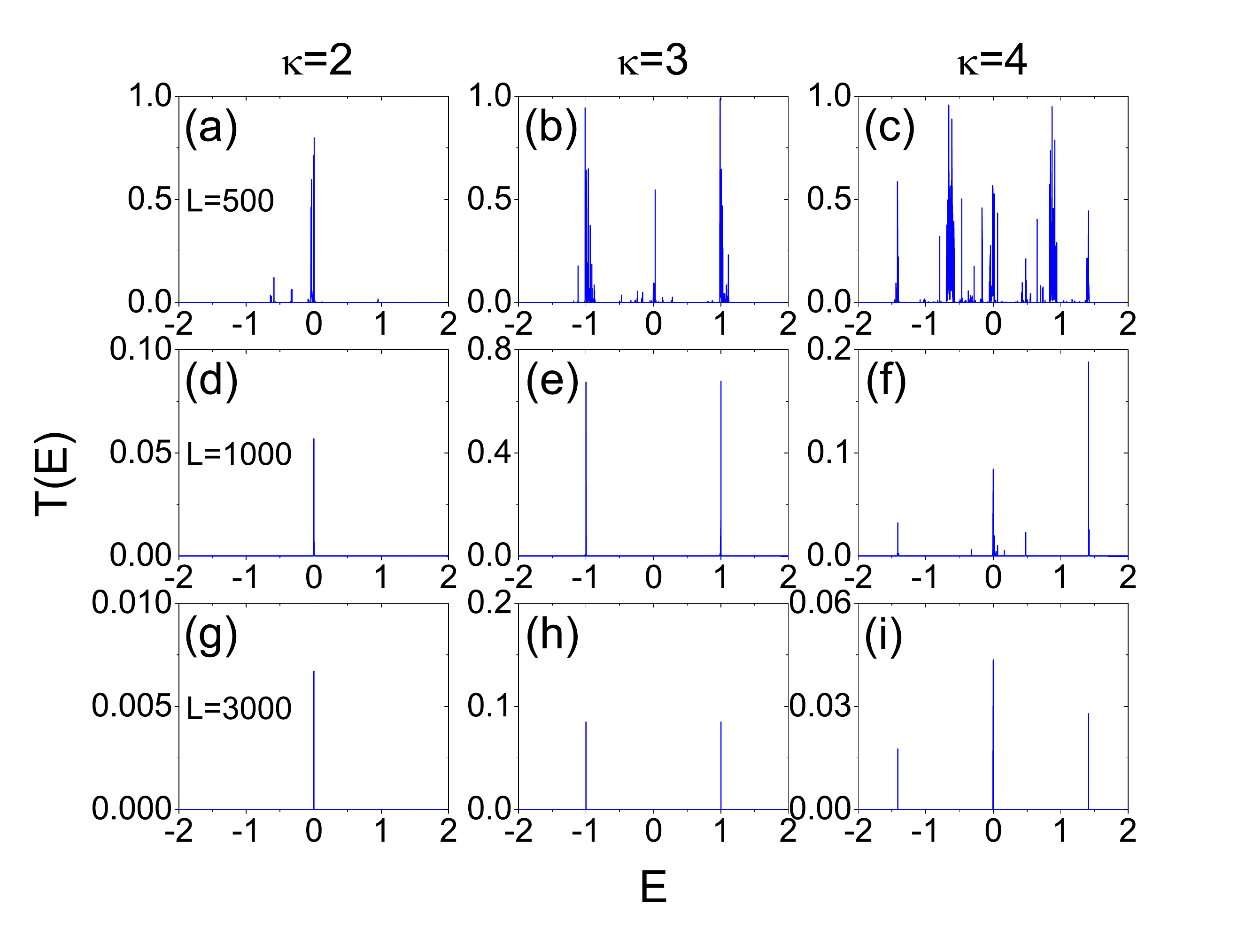}
\caption{Transmittance $T$ plotted versus energy $E$ when $\kappa=2$, 3, 4, $L=500$, 1000, 3000, $W=1$, and $V_0=0$ for a single disorder
configuration. When $L$ is sufficiently large ($L=3000$), $T$ remains non-negligible only at the $(\kappa-1)$ special energy values, such as $E=0$ for $\kappa=2$, $E=\pm1$ for $\kappa=3$, and $E=0$, $\pm\sqrt{2}$ for $\kappa=4$.}
\label{fig1}
\end{figure}

In the $\kappa=1$ case corresponding to the ordinary Anderson model, the transmittance decreases exponentially with $L$ and vanishes in the  $L\rightarrow\infty$ limit for all states \cite{Lif, She}. When the mosaic modulation is turned on, however, the behavior changes substantially.
In figure~\ref{fig1}, we plot the transmittance $T$ as a function of $E$ when $\kappa=2$, 3, 4 and $L=500$, 1000, 3000 for a single disorder configuration.
The parameters $W$ and $V_0$ are fixed to $W=1$ and $V_{0}=0$. When $L$ is small,
$T$ is substantially large for many different $E$ values as shown in figures~\ref{fig1}(a), \ref{fig1}(b), and \ref{fig1}(c). However, when $L$ is sufficiently large (e.g., $L=3000$), we find that $T$ remains non-negligible only at the $(\kappa-1)$ special energy values, such as $E=0$ for $\kappa=2$, $E=\pm1$ for $\kappa=3$, and $E=0$, $\pm\sqrt{2}$ for $\kappa=4$, as shown in figures~\ref{fig1}(g), \ref{fig1}(h), and \ref{fig1}(i). At these energies, the transmittance decays much
more slowly than at other values of $E$ where it decays exponentially with $L$. Recently, we have investigated the characteristics of these spectral points, primarily by analyzing the time-dependent reflectance of the incident wave packet through numerical calculations. \cite{Ngu0}. We have demonstrated that in the long-time limit, almost all the incident wave packets exhibit the exponential localization behavior, whereas those at a discrete set of the $(\kappa-1)$ energy values given by
\begin{eqnarray}
E_{R}=V_{0}+2\cos\left(\frac{\pi}{\kappa}n\right) \quad & (n=1,2,\cdots,\kappa-1)
\label{equation8}
\end{eqnarray}
behave distinctly. We have argued that this is a kind of quasiresonance phenomenon rather than a true resonance and the transmittance at $E_R$ is always less than 1 in our previous work \cite{Ngu0}.

\begin{figure}
\centering
\includegraphics[width=7cm]{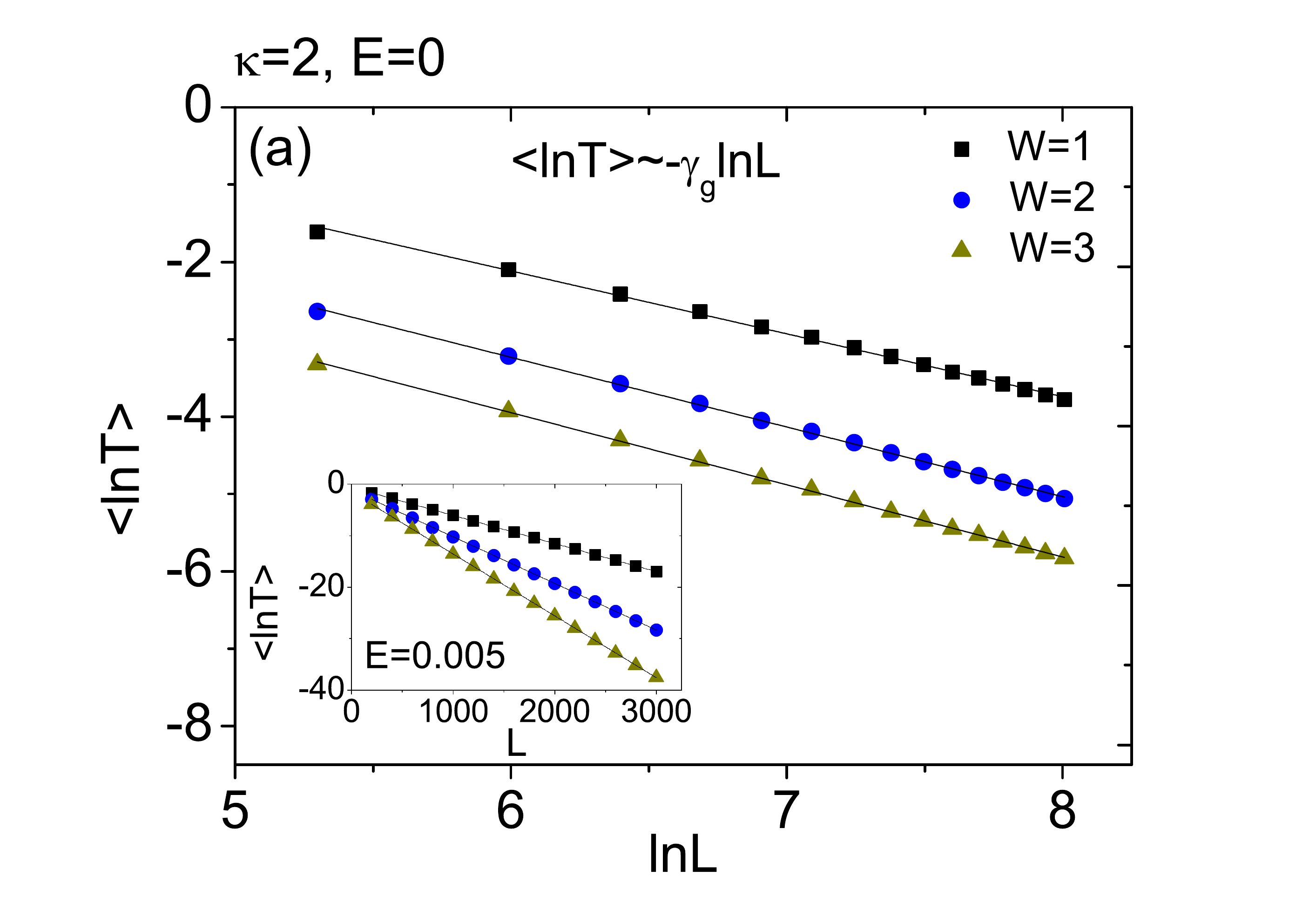}
\includegraphics[width=7cm]{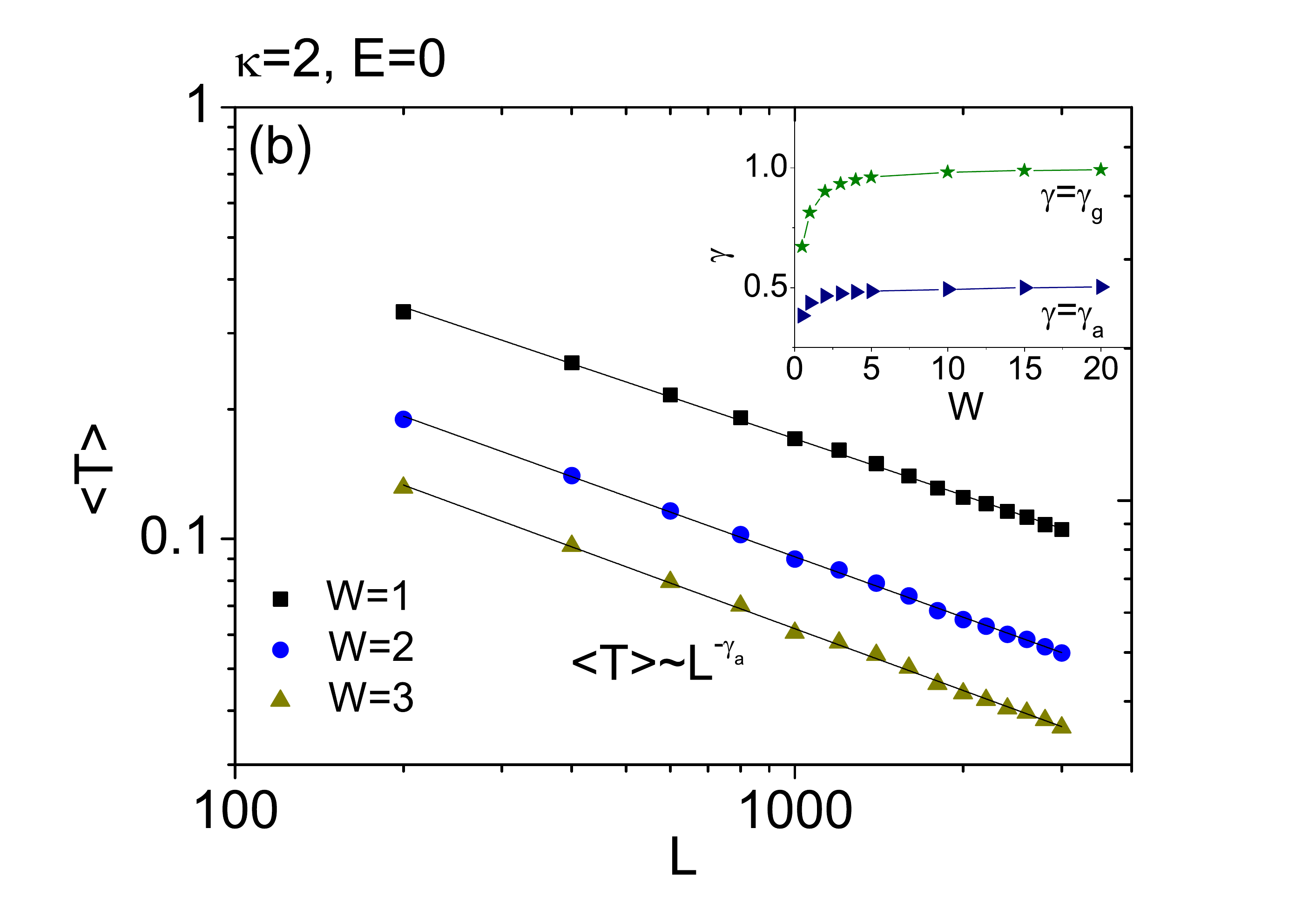}
\includegraphics[width=7cm]{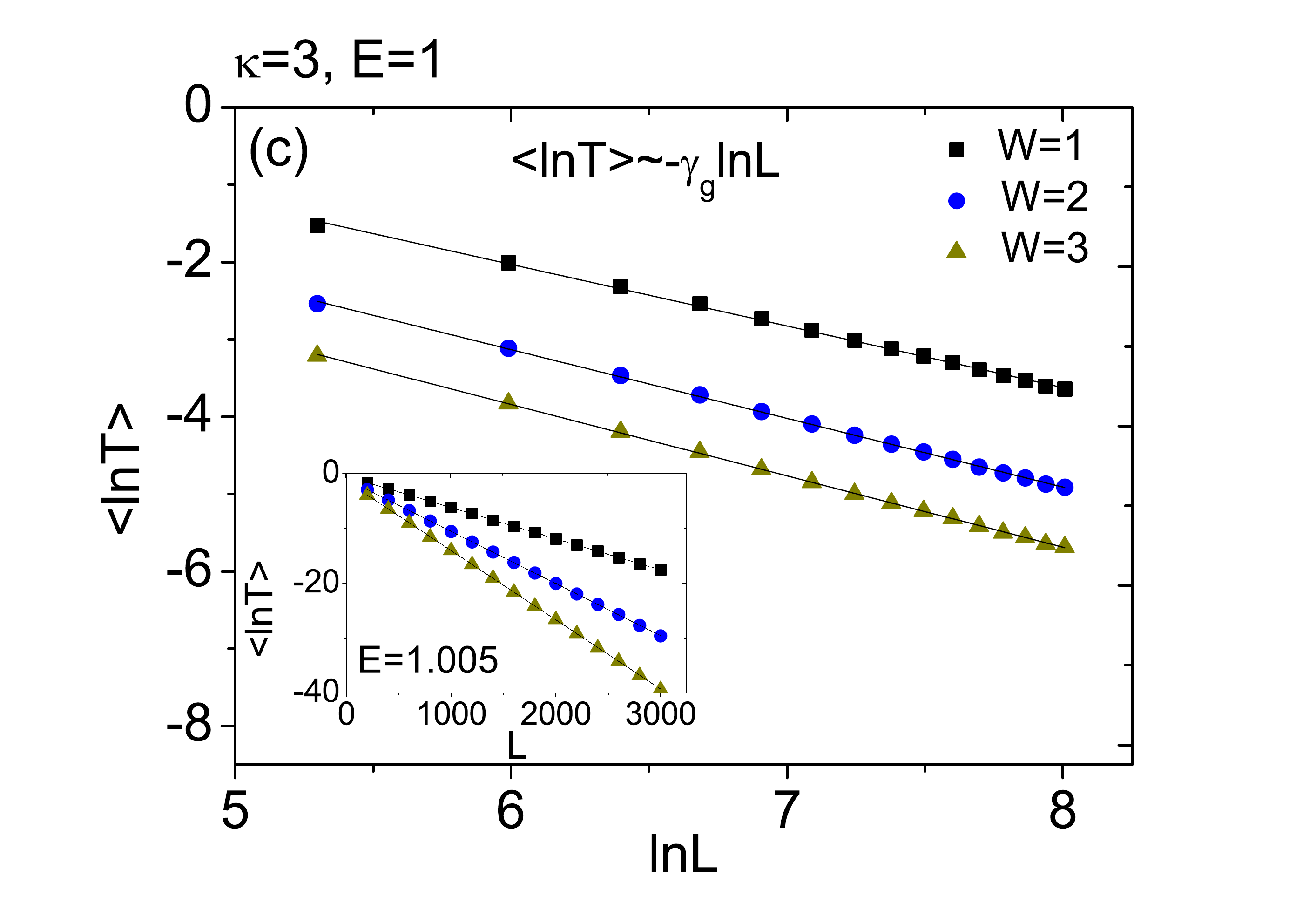}
\includegraphics[width=7cm]{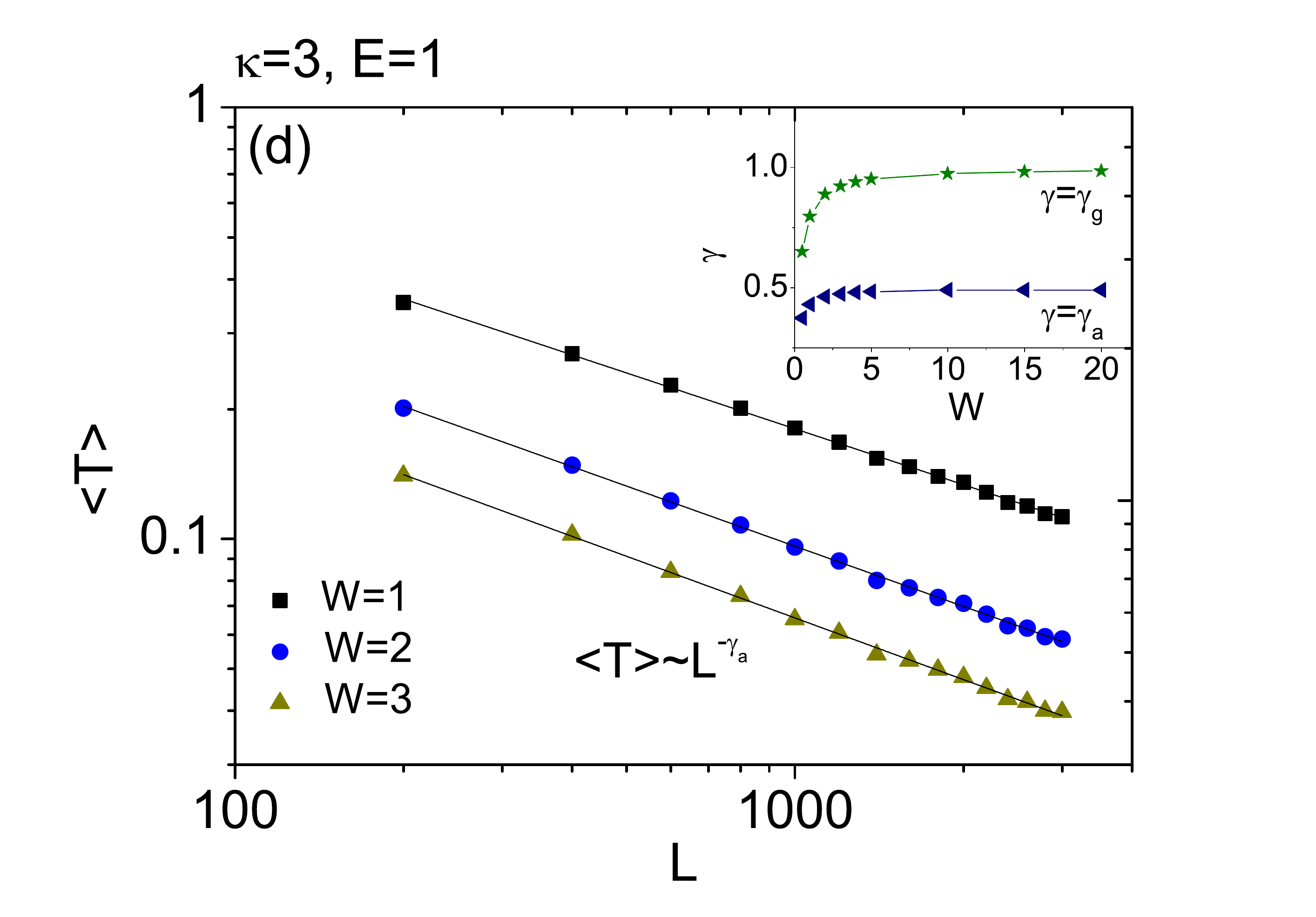}
\includegraphics[width=7cm]{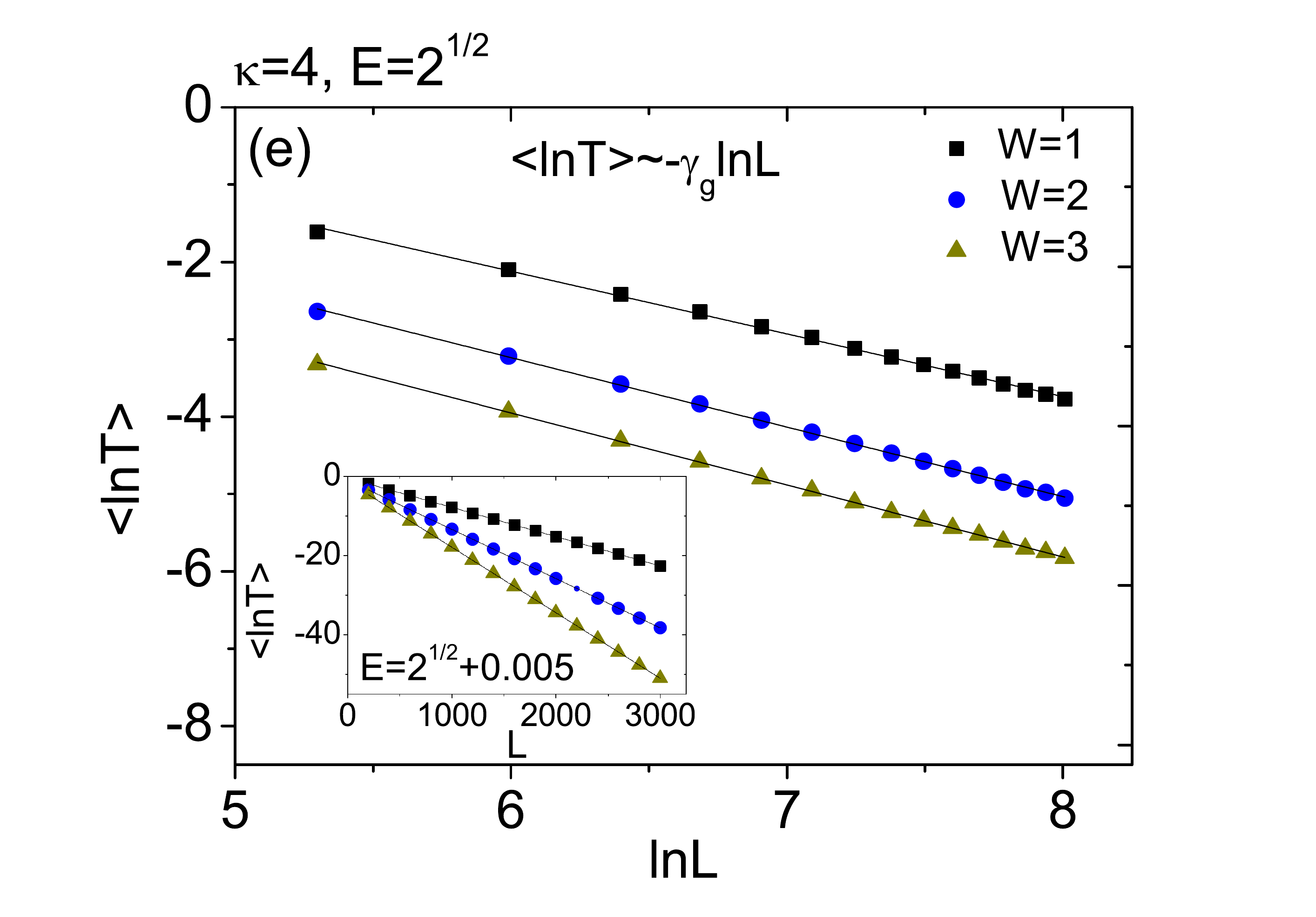}
\includegraphics[width=7cm]{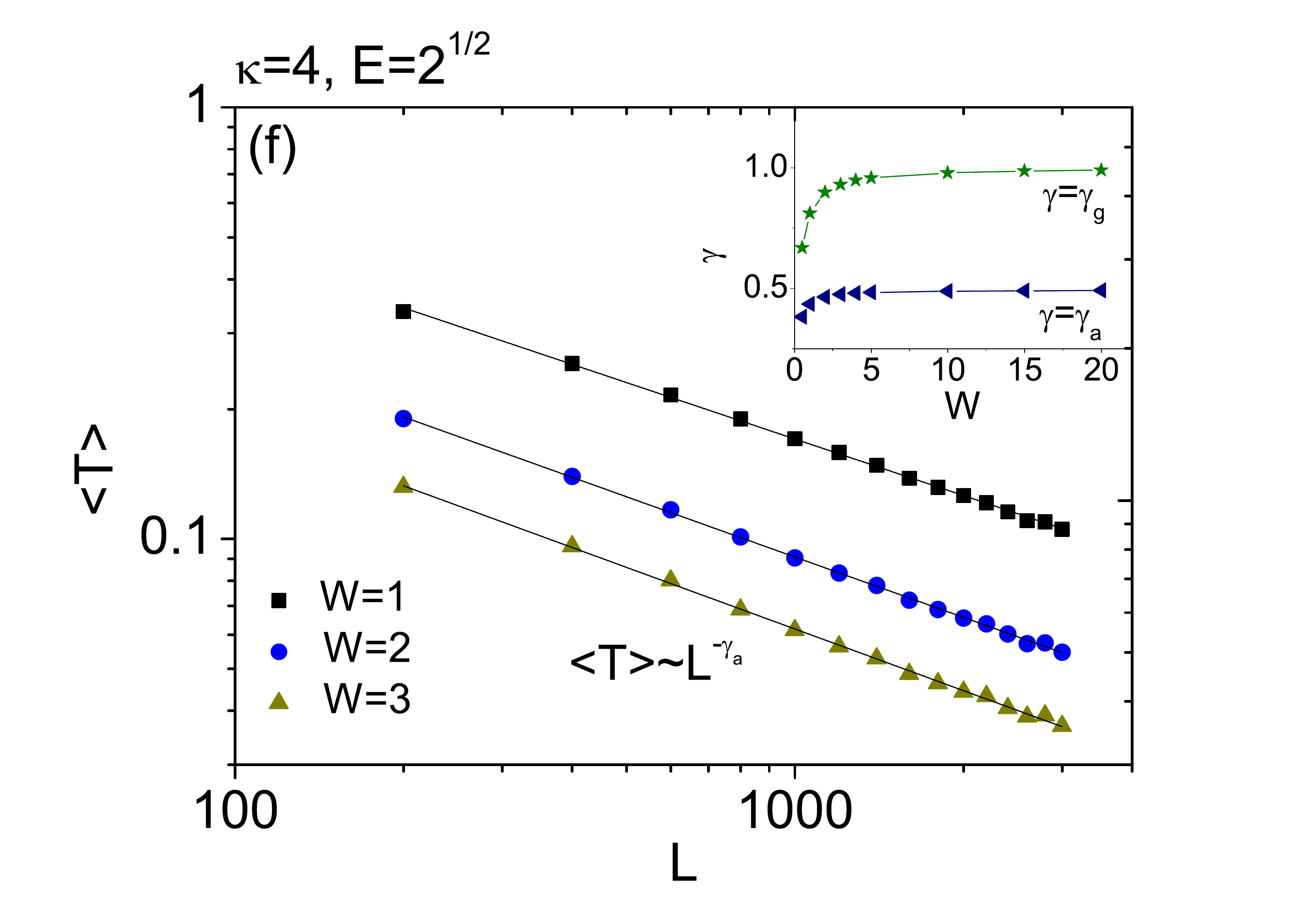}
\caption{
$\langle \ln T \rangle$ plotted versus $\ln L$ and $\langle T \rangle$ plotted versus $L$ on a log-log scale
at the quasiresonance energies (a,b) $E=0$ for $\kappa=2$, (c,d) $E=1$ for $\kappa=3$, and (e,f) $E=\sqrt{2}$ for $\kappa=4$, when $W=1$, 2, 3 and $V_0=0$.
It is demonstrated that $\langle \ln T \rangle\approx -\gamma_g\ln L$ and $\langle T \rangle\propto L^{-\gamma_a}$
in the large-$L$ region. The straight lines represent the data fitting from which the values of the power-law exponents, $\gamma_{a}$ and $\gamma_g$, are extracted.
$\gamma_{a}$ and $\gamma_g$ approach saturation values of $\gamma_{a}\sim 0.5$ and $\gamma_{g}\sim 1$ as $W$ tends to infinity, as illustrated in the right-hand panel insets.
When the energy deviates slightly from the quasiresonance values, exponential localization behavior is observed, where $\langle \ln T \rangle$ behaves as $\langle \ln T \rangle\propto -L$, as depicted in the insets of the left-hand panels.}
\label{fig2}
\end{figure}

In order to get a more detailed understanding of the transmission properties at the quasiresonance energies, we perform a finite-size scaling analysis of the
disorder-averaged quantities $\langle \ln T\rangle$ and $\langle T\rangle$. In the Anderson localized regime, it is expected that $\langle \ln T \rangle\approx -L/\xi_g$ and $\langle T\rangle\propto e^{-L/\xi_a}$ as $L\rightarrow\infty$, where the quantity $\xi_g$ is
conventionally defined to be the localization length.
At the quasiresonance energies, however, markedly different
behaviors are obtained. In figure~\ref{fig2}, we plot $\langle \ln T \rangle$ versus $\ln L$ and $\langle T \rangle$ versus $L$ on a log-log scale at three quasiresonance energies
$E=0$ for $\kappa=2$, $E=1$ for $\kappa=3$, and $E=\sqrt{2}$ for $\kappa=4$, when $W=1$, 2, 3 and $V_0=0$.
Disorder averaging is conducted over 50000 independent disorder configurations, which have been chosen to effectively minimize random fluctuations in the curves at quasiresonance energies.
The results demonstrate that in the large-$L$ region, $\langle \ln T \rangle$ follows $\langle \ln T \rangle \approx -\gamma_g \ln L$, and $\langle T \rangle$ exhibits a power-law behavior of $\langle T \rangle \propto L^{-\gamma_a}$. The straight lines in the plots represent the least-squares fitting, from which we extract the values of the power-law exponents, $\gamma_{g}$ and $\gamma_a$.
$\gamma_{g}$ is obtained from the slope of the curve depicting $\langle \ln T \rangle$ versus $\ln L$ on the linear scale, whereas $\gamma_{a}$ is derived from the slope of the curve showing $\langle T \rangle$ versus $L$ on the log-log scale.
For all the curves, the adjusted R-squared values for the data fitting exceed 0.99.
As $W$ approaches infinity, the exponents saturate at the values $\gamma_{a}\sim 0.5$ and $\gamma_{g}\sim 1$, as shown in the insets of figures~\ref{fig2}(b), \ref{fig2}(d), and \ref{fig2}(f).
In other words, $\langle \ln T \rangle$ and $\langle T \rangle$ exhibit power-law scaling behaviors, as given by
$\langle \ln T \rangle\approx \ln L^{-1/2}$ and $\langle T \rangle\propto L^{-1}$, in the large-$L$
region in the strong disorder limit.
When the energy differs from the quasiresonance values, one can expect exponential localization behavior, such as $\langle \ln T \rangle \propto -L/\xi_g$.

A similar power-law-type behavior of $\langle \ln T\rangle$ and $\langle T\rangle$ occurs in other random systems, which include
1D disordered systems in an external electric (or bias) field and some random models with Kerr-type nonlinearity \cite{Cros1,Sou,Dou,Sha,Iom}.
This behavior is often referred to as
anomalous localization.
It has been argued that the exponent $\gamma_g$ is substantially
larger than $\gamma_a$ due to large fluctuation effects associated with the statistical distribution \cite{Cros1}.
In order to show that the anomalous localization behavior occurs only at the quasiresonance energies, we plot $\langle \ln T \rangle$ versus $L$ for some values of $E$ that deviate only slightly from the quasiresonance energies, in the insets of figures~\ref{fig2}(a), \ref{fig2}(c), and \ref{fig2}(e). In all the cases presented, the transmittance clearly exhibits an exponential decay with $L$, indicating the standard localization behavior. We have also verified numerically that this behavior occurs for all values of $E$ that are not very close to the quasiresonance energies.

\begin{figure}
\centering
\includegraphics[width=12cm]{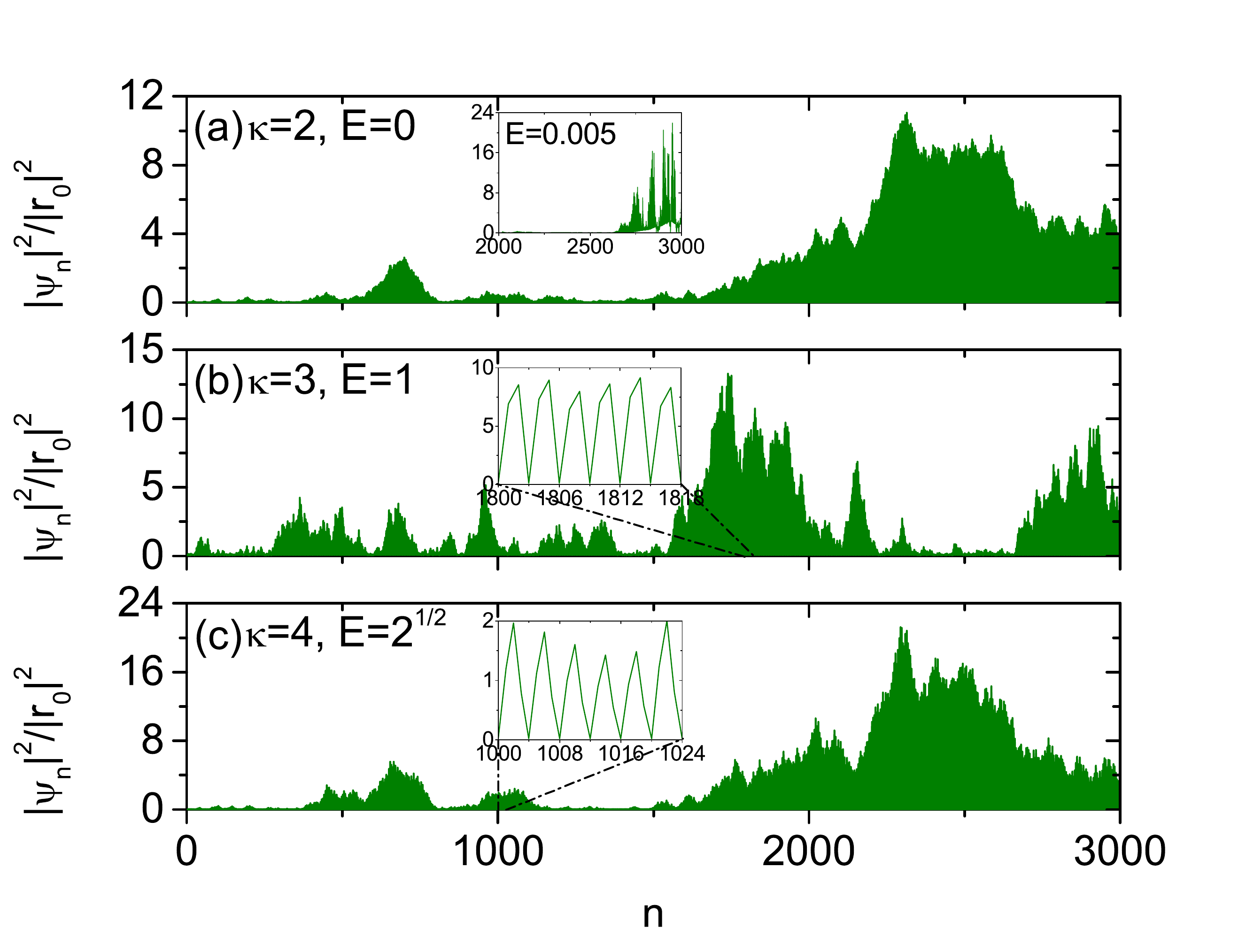}
\caption{Spatial distribution of the intensity of the wave function normalized by that of the incident wave for one particular realization of the random potential, when
a plane wave is incident from the right side. Quasiresonance energies (a) $E=0$ for $\kappa=2$, (b) $E=1$ for $\kappa=3$, and (c) $E=\sqrt{2}$ for $\kappa=4$ are considered. In the inset of (a),
an example of a localized wave function at a slightly different energy
is shown for the comparison with the quasiresonant case. In the insets of (b) and (c), it is shown that the wave functions at the quasiresonance energies have a node structure such that
they vanish at all the sites $n=m\kappa$ for any integer $m$.}
\label{fig3}
\end{figure}

In figure~\ref{fig3}, we show examples of the spatial distribution of wave function intensity at certain quasiresonance energies for a single typical disorder realization, when the system size $L$ is 3000 and a plane wave is incident from the region where $L > 3000$. The wave function intensity is normalized by that of the incident wave, $\vert r_{0}\vert^2$.
We observe that, despite extending across almost the entire system, the wave function displays significant fluctuations in its envelope at quasiresonance energies. This indicates that the associated states are not entirely extended or exponentially localized. Instead, the wave function appears as multiple disconnected patches, a unique feature commonly seen in multifractal or critical states \cite{Cast}. This is in contrast to the wave function corresponding to ordinary localization, as shown in the inset of figure~\ref{fig3}(a).
Investigating the algebraic decay of wave functions in space, as expected for critical states, and conducting a multifractal analysis of the wave function, present intriguing avenues for future research.
In Ref.~\cite{Ngu0}, we have given an argument that at the quasiresonance energies, the wave function should exhibit a node structure where
it is zero at all the lattice sites $m\kappa$ for any integer $m$, as demonstrated in
the insets of figures~\ref{fig3}(b) and \ref{fig3}(c).
A similar intrinsic node structure of the wave function is observed to occur at all the other quasiresonance energies. This is a crucial feature that connects the periodic and disordered mosaic lattice models and gives rise to the critical states at the quasiresonance energies.
\subsection{Participation ratio}
\label{sec32}

In this subsection, we examine the eigenvalue problem for the stationary discrete Schr\"odinger equation given by equation~(\ref{equation2}). We numerically solve the eigenvalue problem of the form
\begin{eqnarray}
{\cal{H}}\Psi=E\Psi
\label{equation9}
\end{eqnarray}
to obtain the eigenvalues $E$ and the corresponding eigenfunctions $\Psi=(\psi_{1},\psi_{2},\dots,\psi_{L})^{\rm T}$. Using the open boundary condition, $\psi_{0}=\psi_{L+1}=0$, the matrix representation of $\cal{H}$ is given by

\begin{eqnarray}
\cal{H} = \left( \begin{array}{cccccc}
			V_{1}		&1				&0		&\cdots 		&0		&0\\
			1	&			V_{2}	&1		&\cdots 		&0		&0\\
           0      &   1    &      V_{3}    & \cdots     &    0   & 0\\
			\vdots 			&\vdots 			&\vdots 	&\cdots 		&\vdots 	&\vdots\\
			0				&0				&0		&\cdots 		&1		&V_{L}
\end{array} \right).
\end{eqnarray}
\label{equation10}

In order to estimate the degree of spatial extension or localization of the eigenstates in disordered systems, we calculate the
participation ratio $P$. For the $k$-th eigenstate $(\psi_{1}^{(k)},\psi_{2}^{(k)},\dots,\psi_{L}^{(k)})^{\rm T}$ with the corresponding eigenvalue $E_{k}$, the participation ratio $P(E_{k})$ is defined by
\begin{eqnarray}
P(E_{k})=\frac{\left(\sum_{n=1}^{L}\left|\psi_{n}^{(k)}\right|^2\right )^2}{\sum_{n=1}^{L}\left|\psi_{n}^{(k)}\right|^4}.
\label{equation11}
\end{eqnarray}
For a finite system, the value of $P$ gives approximately the number of sites over which the $k$-th eigenfunction is extended. It has been known that in the localized case, $P$ is closely related to the localization length, $\xi_g$ \cite{Krim}. The participation ratio generally represents an upper bound for the localization length ($P\geq\xi_g$). When $L$ is sufficiently large, one finds a power-law scaling behavior of the form
\begin{eqnarray}
\langle P(E) \rangle \propto L^{\beta}
\label{equation12}
\end{eqnarray}
with the scaling exponent $\beta$. $\langle P(E) \rangle$ is a double-averaged quantity, where $P(E)$ is obtained by averaging over all eigenstates
within a narrow interval $\Delta E$ ($= 0.1$) around $E$ and $\langle\cdots\rangle$ denotes averaging over a large number of independent random realizations. It is well-established that for extended states, $\langle P(E) \rangle$ increases linearly with $L$ and the exponent $\beta$
is equal to 1. In contrast, for exponentially localized states, $\beta$ is zero and $\langle P(E) \rangle$ does not depend on $L$ and converges to a constant value in the $L\to \infty$ limit. For critical states, the scaling exponent should be in the range of $0<\beta<1$. Therefore the finite-size scaling analysis of $\langle P(E) \rangle$ provides very useful information about the nature of the states.

\begin{figure}
\centering
\includegraphics[width=12cm]{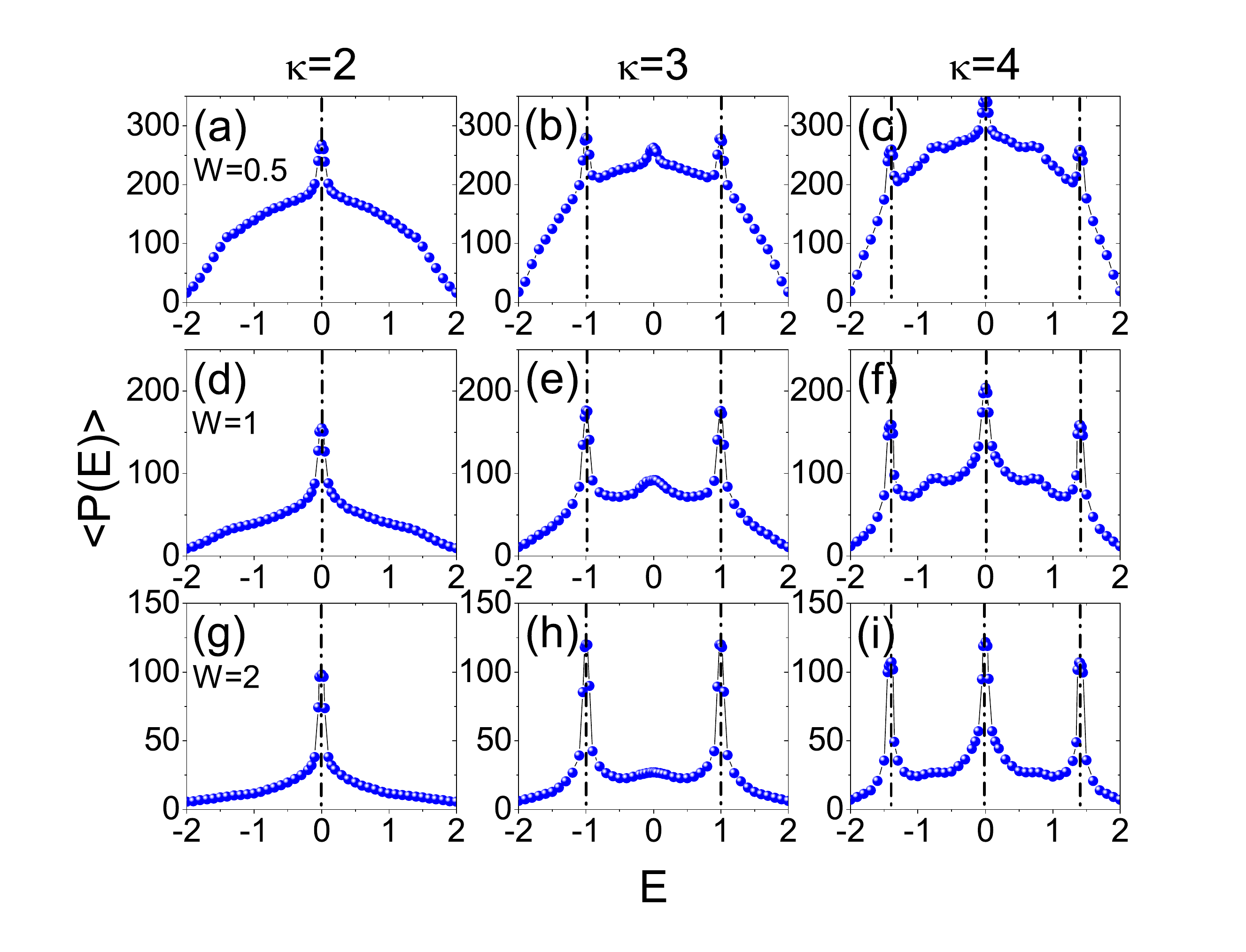}
\caption{Participation ratio averaged over 1000 distinct disorder configurations, $\langle P(E) \rangle$,  plotted versus energy $E$ when $\kappa=2$, 3, 4, $W=0.5$, 1, 2, $V_0=0$,
and $L=1000$.
$\langle P(E) \rangle$ is sharply enhanced at the quasiresonance energy values $E=0$ for $\kappa=2$, $E=\pm 1$ for $\kappa=3$, and $E=0,~\pm\sqrt{2}$ for $\kappa=4$.}
\label{fig4}
\end{figure}

\begin{figure}
\centering
\includegraphics[width=11cm]{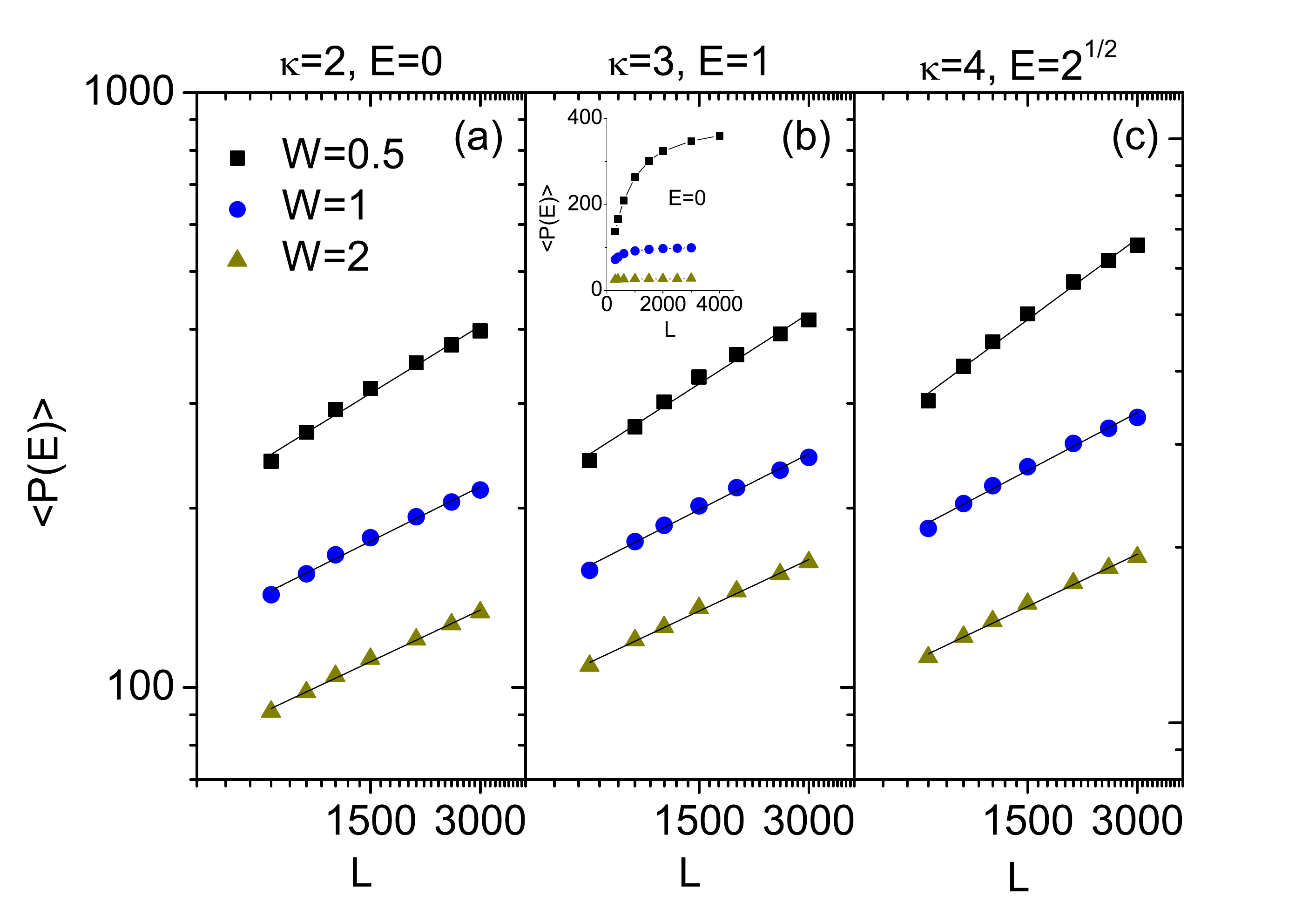}
\caption{Disorder-averaged participation ratio $\langle P(E) \rangle$ plotted versus $L$ on a log-log scale
at (a) $E=0$ for $\kappa=2$, (b) $E=1$ for $\kappa=3$, and (c) $E=\sqrt{2}$ for $\kappa=4$, when $W=0.5$, 1, 2 and $V_0=0$.
In all cases, $\langle P(E) \rangle$ scales as $L^{\beta}$ with $0<\beta<1$ for sufficiently large $L$.
The inset in (b) displays $\langle P(E) \rangle$ versus $L$ on a linear scale when $E=0$ and $\kappa=3$. }
\label{fig5}
\end{figure}

\begin{figure}
\centering
\includegraphics[width=11cm]{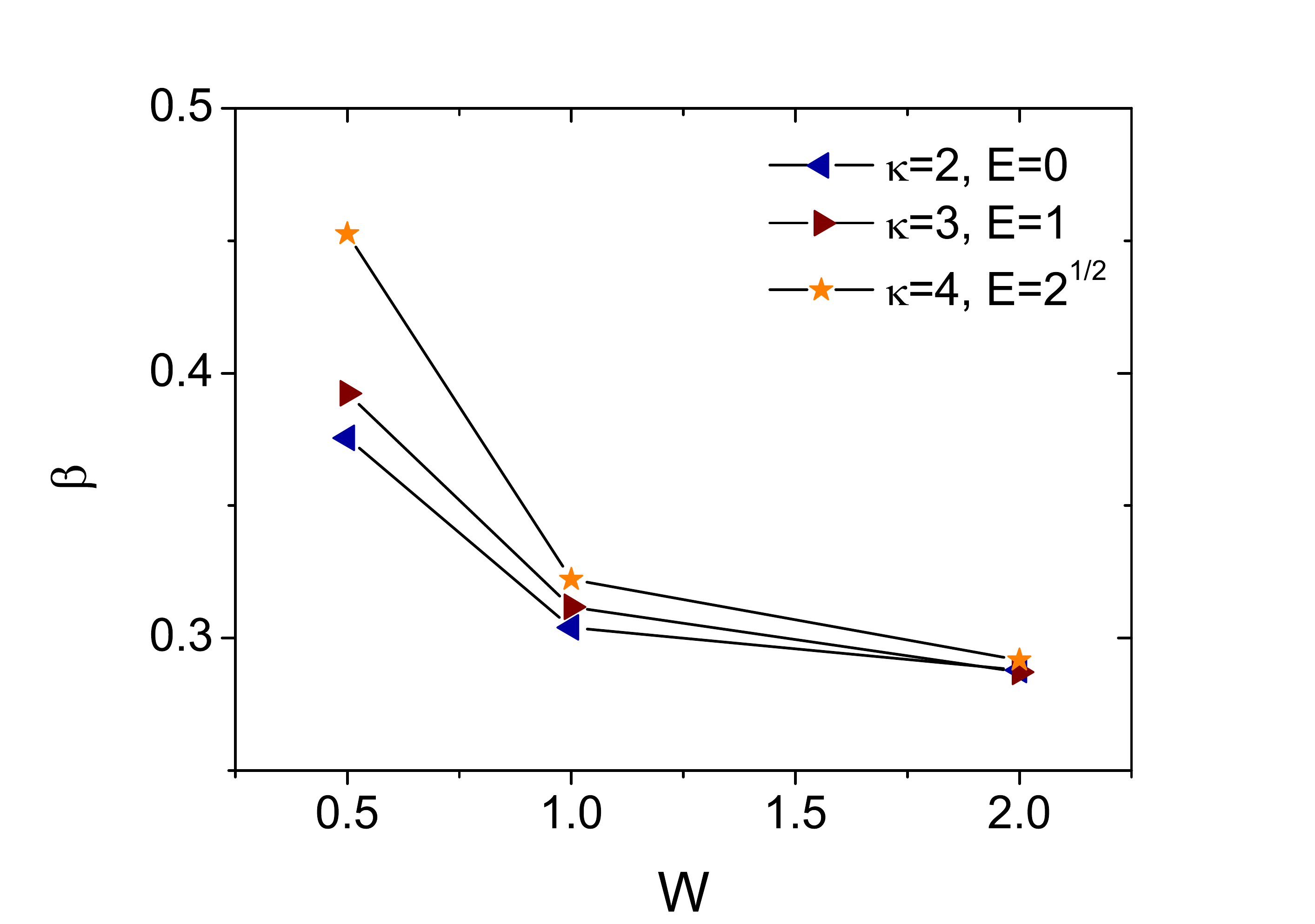}
\caption{
Dependence of the scaling exponent $\beta$ at quasiresonance energies $E=0$ for $\kappa=2$, $E=1$ for $\kappa=3$, and  $E=\sqrt{2}$ for $\kappa=4$ on the disorder strength $W$.
$\beta$ initially decreases and then saturates to the same values as $W$ increases.}
\label{fig44}
\end{figure}

In figure~\ref{fig4}, we show $\langle P(E) \rangle$ obtained by averaging over 1000 distinct disorder realizations as a function of energy $E$ when $\kappa=2$, 3, 4, $W=0.5$, 1, 2, $V_0=0$, and $L=1000$.
We find that $\langle P(E) \rangle$ is sharply enhanced at all the quasiresonance energy values $E_R$ given by equation~(\ref{equation8}),
which implies that the localization is strongly suppressed at such energies.
These results are fully consistent with those obtained in the previous subsection. The sharp peak at $E=0$ is expected to occur
only when $\kappa$ is even. However, if we pay close attention to the case of $\kappa=3$, a relatively small peak at $E=0$ is clearly seen when the disorder is sufficiently weak (e.g., $W=0.5$) [see figure~\ref{fig4}(b)]. This peak occurs due to the band-center anomaly (or Kappus-Wegner anomaly) and arises from the hidden symmetry at this spectral point \cite{Kapp,Dey,Scho,Ngu3}. One important difference between this peak and those at the quasiresonance energies will be pointed out below.

In order to understand the critical nature of states at quasiresonance energies, we perform a finite-size scaling analysis of $\langle P(E) \rangle$ based on the power-law ansatz, equation~(\ref{equation12}). In figure~\ref{fig5}, we present the results of numerical calculations of $\langle P(E) \rangle$ versus $L$ on a log-log scale at $E=0$ for $\kappa=2$, $E=1$ for $\kappa=3$, and $E=\sqrt{2}$ for $\kappa=4$, when $W=0.5$, 1, 2 and $V_0=0$. In all cases, $\langle P(E) \rangle$ is found to behave as a power law $\langle P(E) \rangle\propto L^{\beta}$ with $0<\beta<1$. The scaling exponent $\beta$ has been determined from the slope of the plot of $\langle P(E) \rangle$ versus $L$ on a log-log scale. The dependence of $\beta$ on the disorder strength $W$ is shown in figure~\ref{fig44}.
We find that $\beta$ is a decreasing function of $W$ and converges to a saturation value in the strong disorder regime. Our numerical results seem to suggest that the limiting value of $\beta$ is approximately independent of the quasiresonance energy. These results indicate that the eigenstates at quasiresonance energies are neither extended nor exponentially localized, but critical states.

We have also performed a finite-size scaling analysis of $\langle P(E) \rangle$ at $E=0$ for $\kappa=3$, which corresponds to
the band-center anomaly. The result shows that the exponent $\beta$ approaches towards zero as $L\rightarrow \infty$ regardless of the disorder strength $W$ \cite{Krim}. This implies that although the participation ratio at the band center increases anomalously, the state at this spectral point is still exponentially localized and is fundamentally different from the quasiresonant states appearing at $E_R$.

\subsection{Time-dependent wave packet dynamics}
\label{sec33}

Up to this point, our emphasis has been on the stationary properties
of excitations in the disordered mosaic lattice model.
Here, we shift our focus to the dynamic characteristics and explore how initially localized wave packets propagate over time. Previous research has demonstrated that critical states typically lead to subdiffusive wave packet spreading in the long-time limit \cite{Hir, Lar}.
While there are various types of initial wave packets that can be employed, including a single-site $\delta$ function wave packet, it is most suitable to use a Gaussian wave packet with a finite initial momentum, as defined by \cite{San}
\begin{eqnarray}
C_{n}\left(t=0\right)=A_\sigma\exp \left[-\frac{\left(n-n_{0}\right)^2}{4\sigma^2}+iq\left(n-n_{0}\right)\right],\nonumber\\
\label{equation13}
\end{eqnarray}
where $\sigma$ and $n_{0}$ are the spatial width and the initial position of the wave packet, respectively. $A_\sigma$ is the normalization constant that has to be determined according to $\sum_{n}\left\vert C_{n}(t)\right\vert^2=1$. It is reminded that $q$ is the wave number related to $E$ by $E=2\cos q$. It is worth mentioning that the initial wave packet defined by equation~(\ref{equation13}) possesses a momentum distribution that is also a Gaussian, centered around  $q$, and its momentum width is inversely proportional to the spatial width $\sigma$. In order to ensure that the lattice chain's dynamic evolution is governed by a clearly defined energy $E$, it is essential to employ wave packets with a narrow momentum distribution. For this reason, the spatial width $\sigma$ is selected to be adequately broad to satisfy the condition that $q\sigma\gg 1$ \cite{Del1}.

We solve numerically the time-dependent discrete Schr\"odinger equation, equation~(\ref{equation1}), to determine the time evolution of an initially localized wave packet given by equation~(\ref{equation13}).
We utilized the Adams-Moulton method to solve ordinary differential equations
and emplyed the IMSL subroutine IVPAG. To characterize the wave packet's
spatial spreading over time, we calculate dynamic quantities such as the mean-square displacement,
spatial probability distribution, participation number, and return probability.

\subsubsection{Mean-square displacement and spatial probability distribution}
\label{sec331}

The first quantity we focus on is the mean-square displacement $m^{2}(t)$ defined by
\begin{eqnarray}
m^{2}(t)=\sum_{n=1}^{L}(n-n_{0})^2\left\vert C_{n}(t)\right\vert^2,
\label{equation14}
\end{eqnarray}
which measures the difference between the wave packet's position at time $t$ and its initial position \cite{Joh}. The asymptotic spreading of the ensemble-averaged quantity $\langle m^2(t) \rangle$ can be fitted using a power-law ansatz
\begin{eqnarray}
\langle m^2(t) \rangle\propto t^{\eta},
\label{equation15}
\end{eqnarray}
where $\langle \cdots \rangle$ denotes the averaging over independent realizations of the disorder. By analyzing the scaling exponent $\eta$, we can identify the various transport regimes, such as ballistic ($\eta=2$), superdiffusive ($1<\eta<2$), diffusive ($\eta=1$), subdiffusive ($0<\eta<1$), and localized ($\eta=0$) regimes.

In order to prevent unwanted boundary effects in the numerical simulations, the system size $L$ must be adequately large such that the wave function amplitude is insignificant at the edges for the longest time considered. Subsequently, the numerical calculations are conducted using a system size of $L=40000$, the longest evolution time $t_{\rm max}=10^4$, and a time step of $\Delta t=0.1$. This choice guarantees that $\left\vert C_{n}(t_{\rm max})\right\vert^2<10^{-50}$ at the edges. In addition, we have chosen $\sigma=20$ and $n_{0}=L/2$ and the value of all calculated quantities is averaged over 50 independent disorder realizations.
Employing a limited number of random configurations in the calculation of time-dependent quantities is essential to keep computation times reasonable.
However, it may lead to the persistence of some random fluctuations in the averaged curves.
This stands in contrast to the use of a large number of configurations in the calculation of time-independent quantities, where calculations are significantly more efficient.
The magnitude of boundary effects will increase if different initial conditions are chosen, such as placing the wave packet closer to one end of the lattice chain than the other. However, when the lattice chain is sufficiently long within the considered time limit $t_{\rm max}$, using different boundary conditions will yield the same results.

\begin{figure}
\centering
\includegraphics[width=9cm]{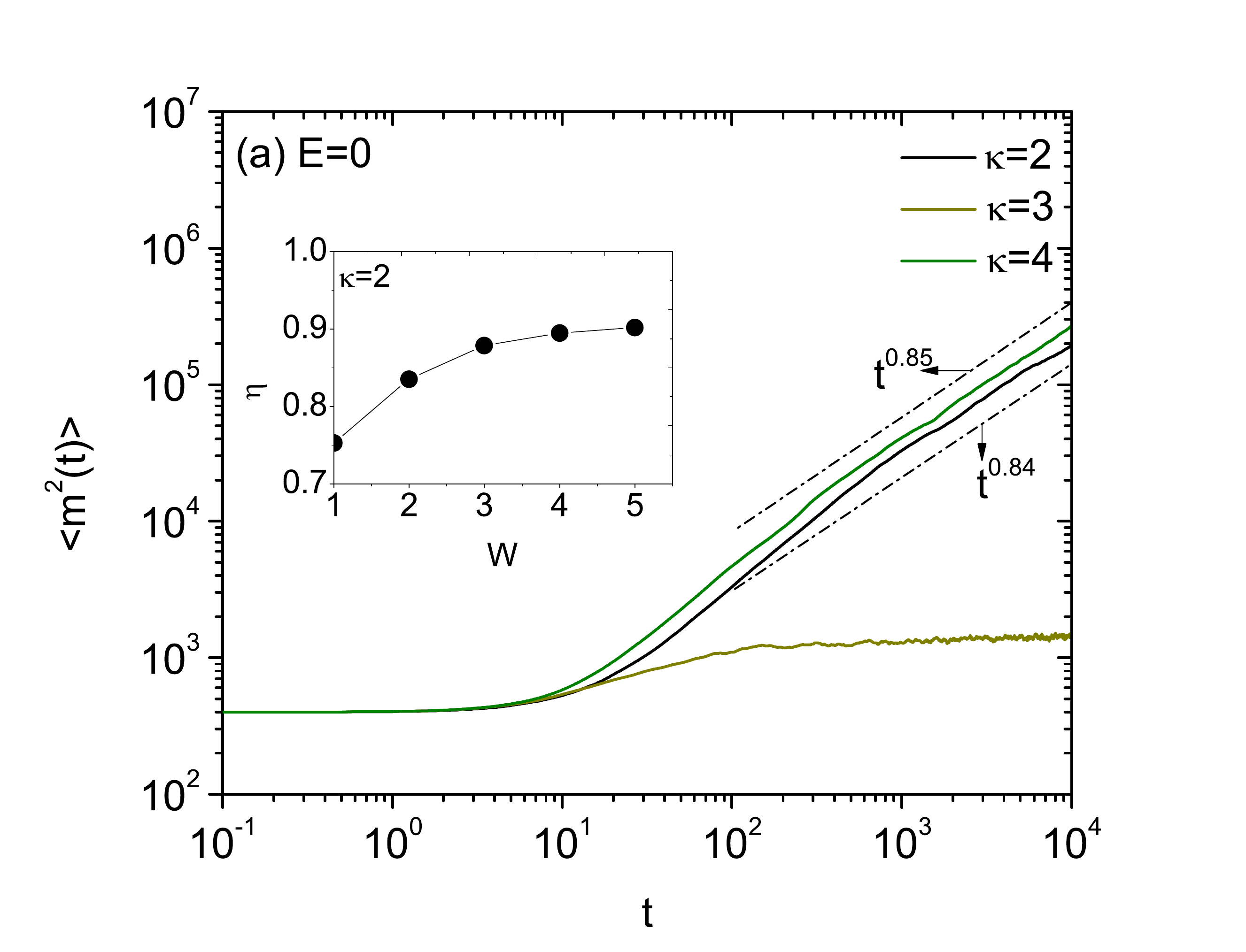}
\includegraphics[width=9cm]{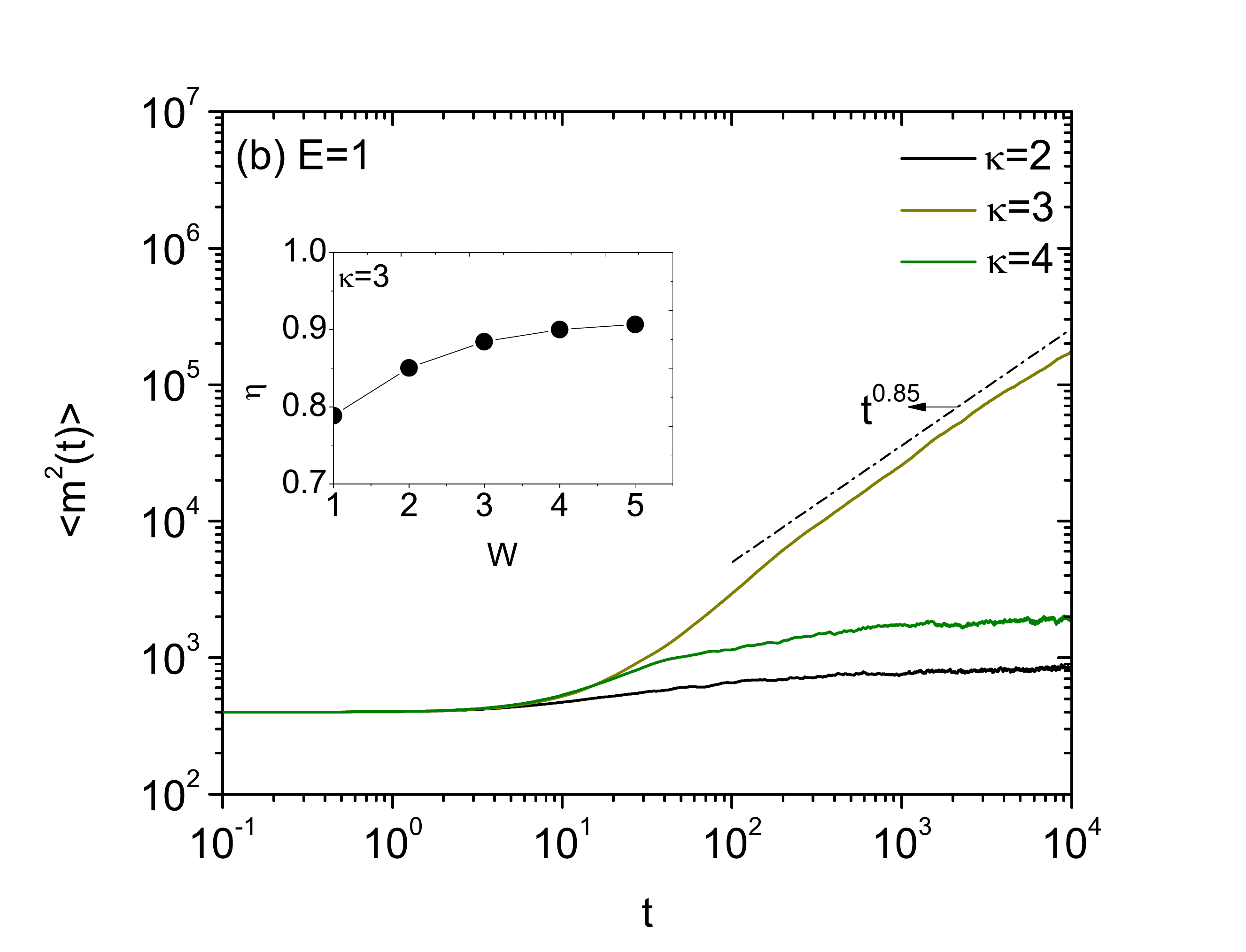}
\includegraphics[width=9cm]{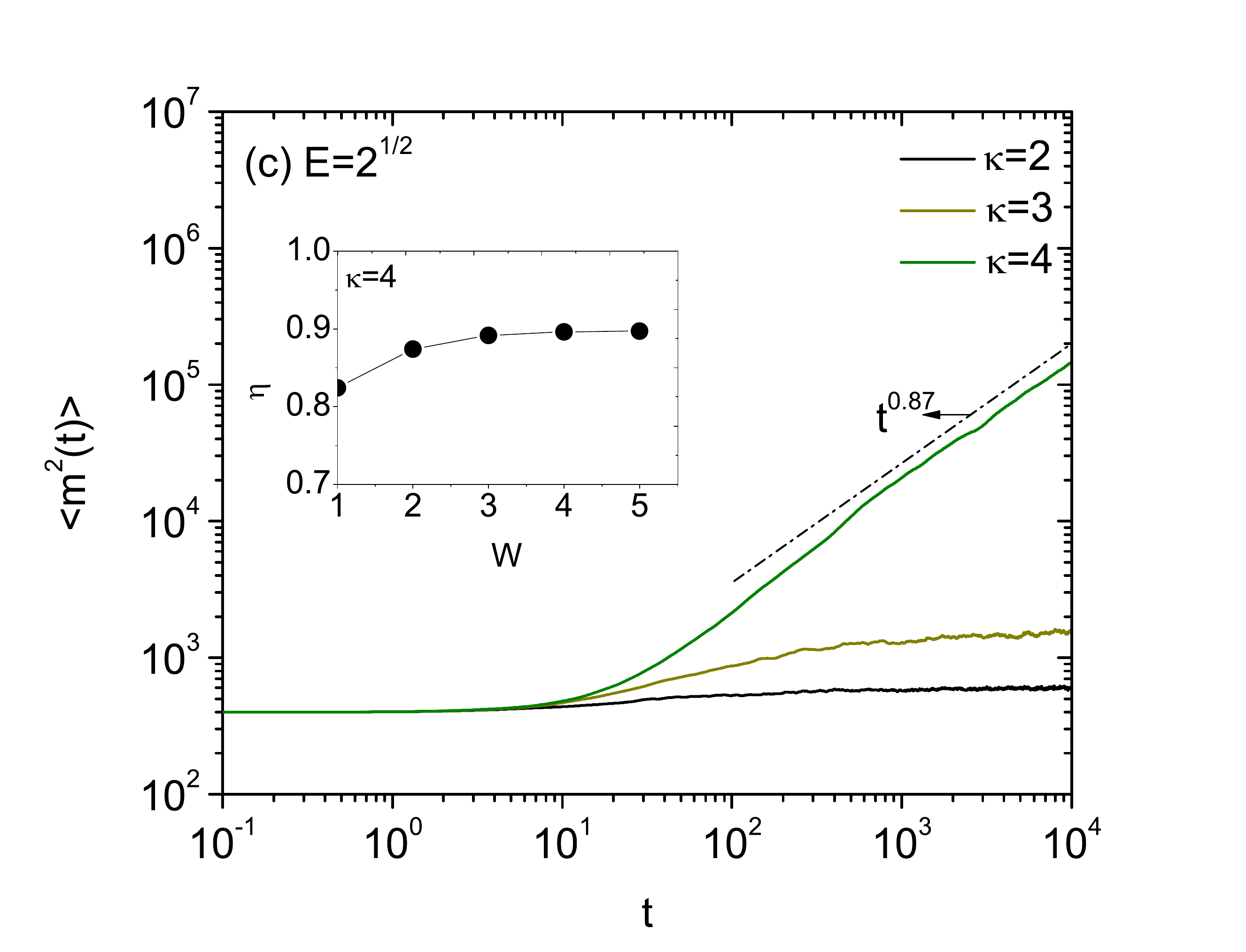}
\caption{Mean-square displacement obtained by averaging over 50 independent disorder realizations, $\langle m^2(t) \rangle$, plotted versus time $t$ at (a) $E=0$, (b) $E=1$, and (c) $E=\sqrt{2}$, when $\kappa=2$, 3, 4, $W=2$, and $V_0=0$. All the results exhibit subdiffusive transport at the quasiresonance energies such that $\langle m^2(t) \rangle\propto t^{\eta}$ with $0<\eta<1$. When the energy is not at the quasiresonance energies, $\langle m^2(t) \rangle$ is seen to converge to a constant value and $\eta$ approaches to 0. The $W$ dependence of the scaling exponent $\eta$ is shown in the insets.}
\label{fig6}
\end{figure}

In figure~\ref{fig6}, we plot the time evolution of $\langle m^2(t) \rangle$ of an initially localized Gaussian wave packet for $\kappa=2$, 3, and 4 at $E=0$, 1, and $\sqrt{2}$, when $W=2$ and $V_0=0$.
We find that when the wave packet is released with an initial velocity that does not satisfy the quasiresonance condition, equation~(\ref{equation8}), the temporal growth of $\langle m^2(t) \rangle$ ceases completely in the long-time limit. This behavior is characterized by the scaling exponent $\eta=0$ in equation~(\ref{equation15}), which indicates the usual Anderson localization. In contrast, when the initial velocity of the wave packet does satisfy the quasiresonance condition, a subdiffusive spreading with $0<\eta<1$ appears at long times. The exponent $\eta$ has been determined by the fitting of the $\langle m^2(t) \rangle$ data in the range of $100\leq t\leq 10^4$. When $W$ is 2, we obtain $\eta\approx 0.84$ for $\kappa=2$ and $\eta\approx 0.85$ for $\kappa=4$ at $E=0$,  $\eta\approx 0.85$ for $\kappa=3$ at $E=1$, and $\eta\approx 0.87$ for $\kappa=4$ at $E=\sqrt{2}$. The $W$ dependence of $\eta$ at these quasiresonance energies is shown in the insets. We find that $\eta$ is an increasing function of $W$, but its value is always less than 1, which indicates that the subdiffusive transport behavior observed at quasiresonance energies remains robust. Like the other power-law exponents $\gamma_a$, $\gamma_{g}$, and $\beta$ calculated in the previous subsections, we note that the exponent $\eta$ approaches the same saturation value at all quasiresonance energies as $W$ increases.

\begin{figure}
\centering
\includegraphics[width=9cm]{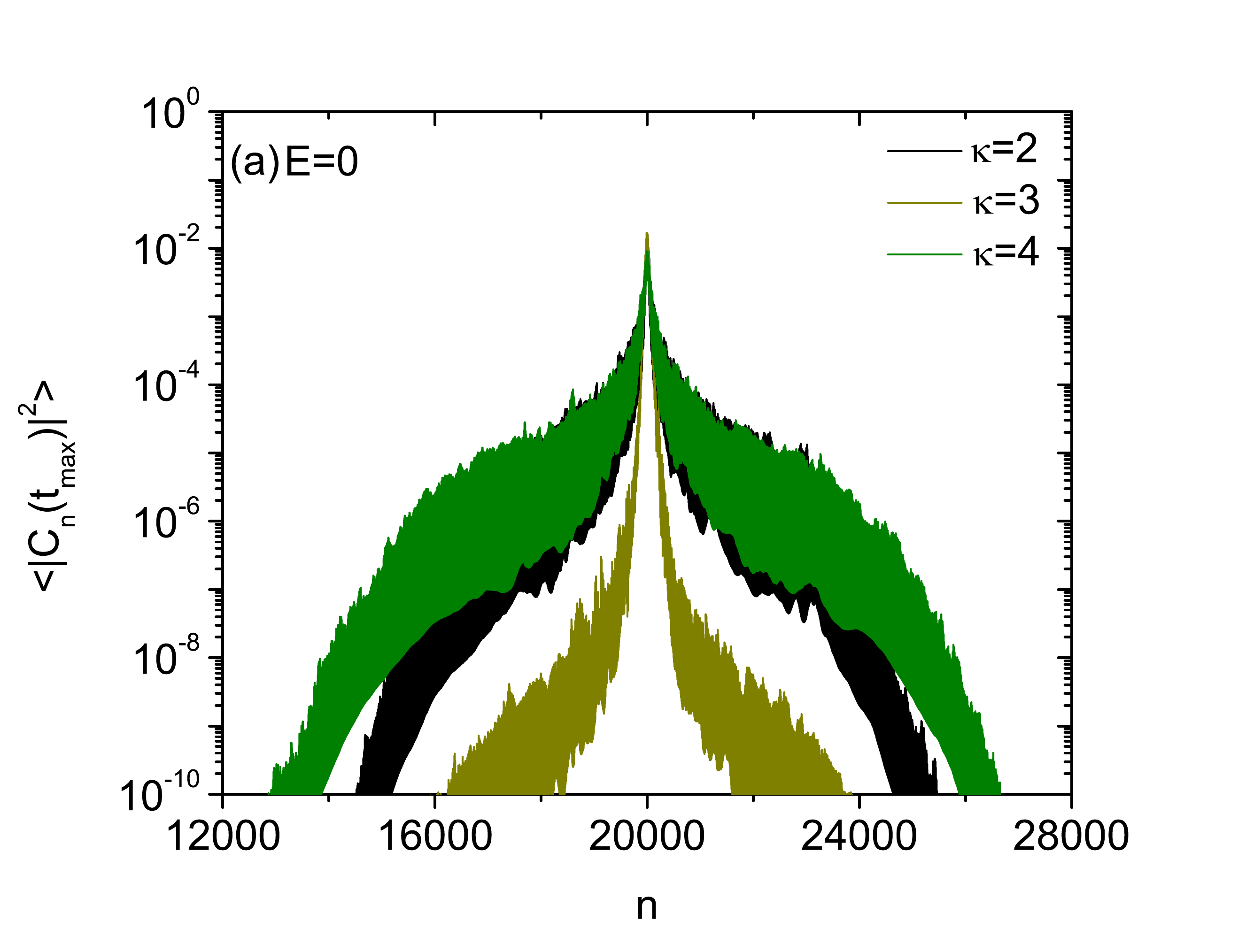}
\includegraphics[width=9cm]{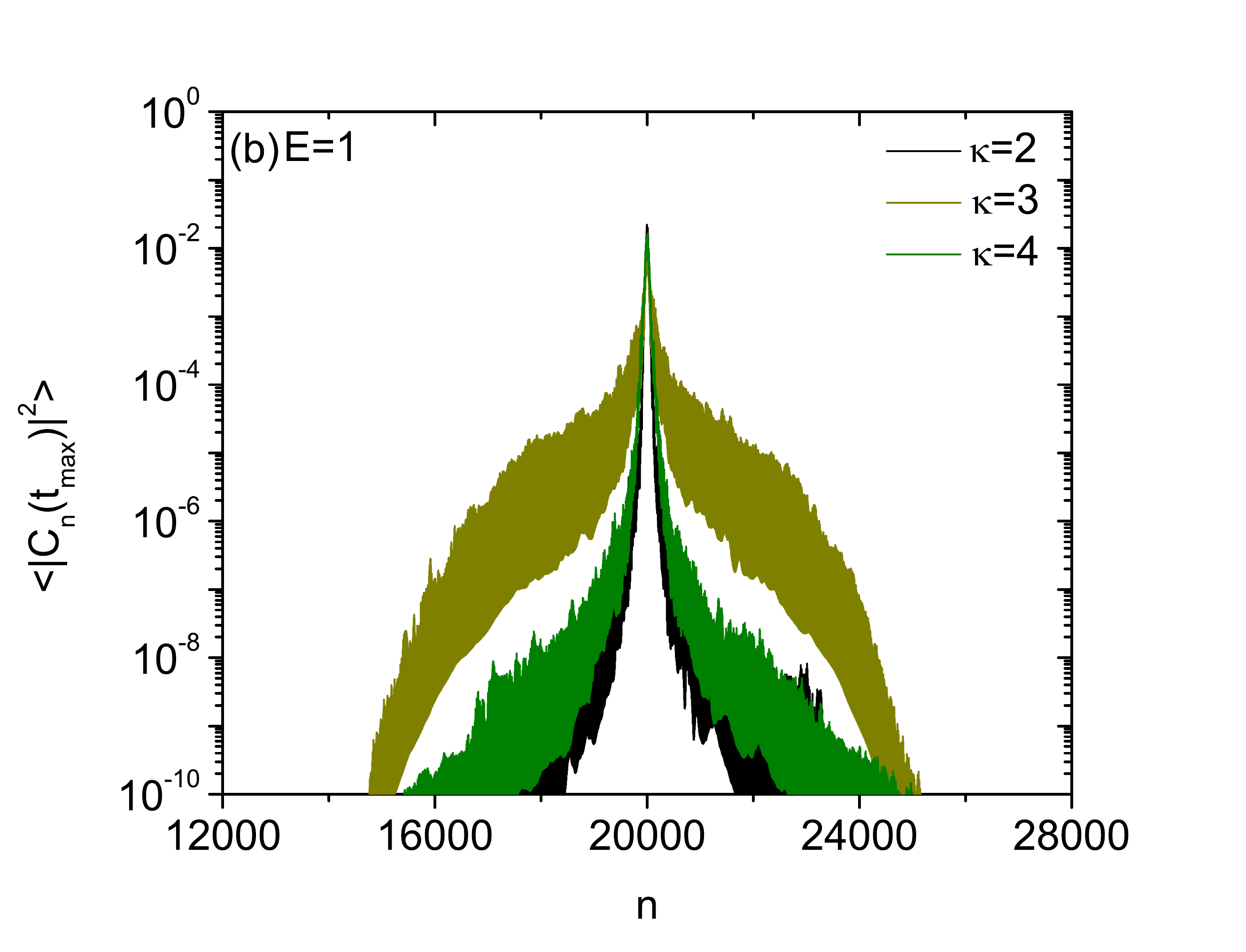}
\includegraphics[width=9cm]{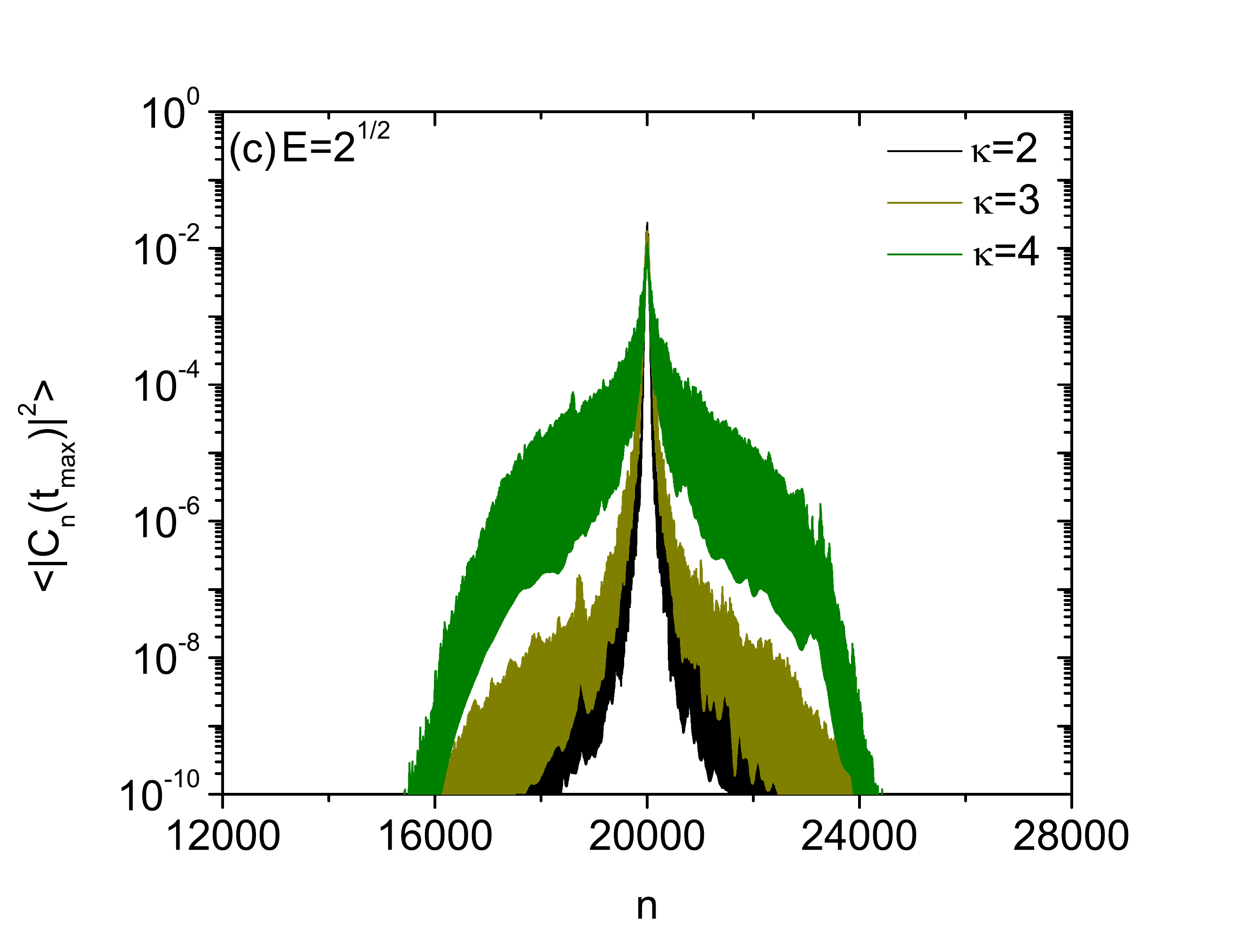}
\caption{Disorder-averaged probability distribution $\langle\left\vert C_{n}(t)\right\vert^2\rangle$ versus site index $n$ at $t=t_{\rm max}=10^4$ when $\kappa=2$, 3, 4, $W=2$, $V_0=0$, and (a) $E=0$, (b) $E=1$, (c) $E=\sqrt{2}$. The values of $\left\vert C_{n}(t)\right\vert^2$ are averaged over the same 50 disorder realizations as in figure \ref{fig6}.}
\label{fig7}
\end{figure}

A more visual picture of the wave packet spreading over time can be obtained from plotting the averaged spatial probability distribution $\langle |C_{n}(t)|^2\rangle$. In figure~\ref{fig7}, we show $\langle |C_{n}(t)|^2\rangle$ versus site index $n$ at $t=t_{\rm max}=10^4$ in a logarithmic plot. The other parameters are the same as in figure~\ref{fig6}.
Observations reveal three significant characteristics. The first feature involves the emergence of two distinct profiles for the averaged field distribution that are reliant on the localization behavior of the states.
When the energy corresponds to quasiresonance energy, such as $E=0$ for $\kappa=2$ and 4, $E=1$ for $\kappa=3$, and $E=\sqrt{2}$ for $\kappa=4$,
the average field pattern displays a Gaussian-like characteristic profile
and the logarithm of $\langle |C_{n}(t)|^2\rangle$ decays parabolically away from the central region of the lattice.
In contrast, when the energy does not correspond to quasiresonance energy,
the average field pattern decays exponentially away from the central region of the lattice and the logarithm of $\langle |C_{n}(t)|^2\rangle$ follows a linear curve, indicating the usual localization behavior.
In previous studies of localization phenomena in nonlinear systems, similar characteristic field profiles have been observed experimentally in synthetic photonic lattices and ultracold-atom systems \cite{Sch, Roa}.
The second feature is that when the quasiresonance condition is satisfied, the width of the wave packet gets broader when its initial velocity is larger. This is consistent with the calculated values of $\langle m^2(t) \rangle$ presented in figure~\ref{fig6}.
The third distinct feature is the emergence of a pronounced peak in the probability distribution at the center of the lattice, even when at quasiresonance energies. This indicates that while the edges of the wave packet undergo subdiffusive expansion within the lattice, its central part will remain localized near its original position for an extended period. This result will be further corroborated below.

\subsubsection{Participation number and return probability}
\label{sec332}

\begin{figure}
\centering
\includegraphics[width=9cm]{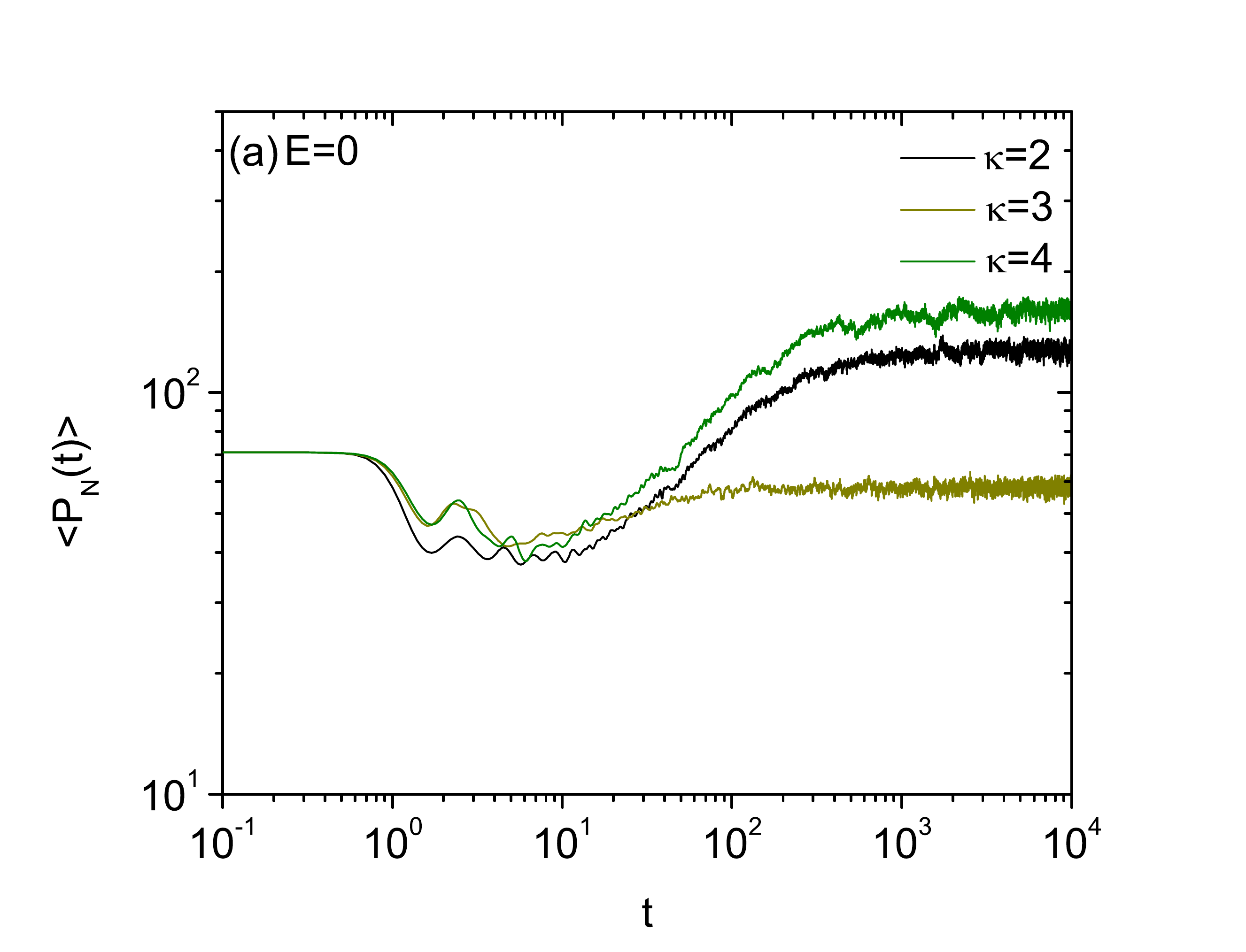}
\includegraphics[width=9cm]{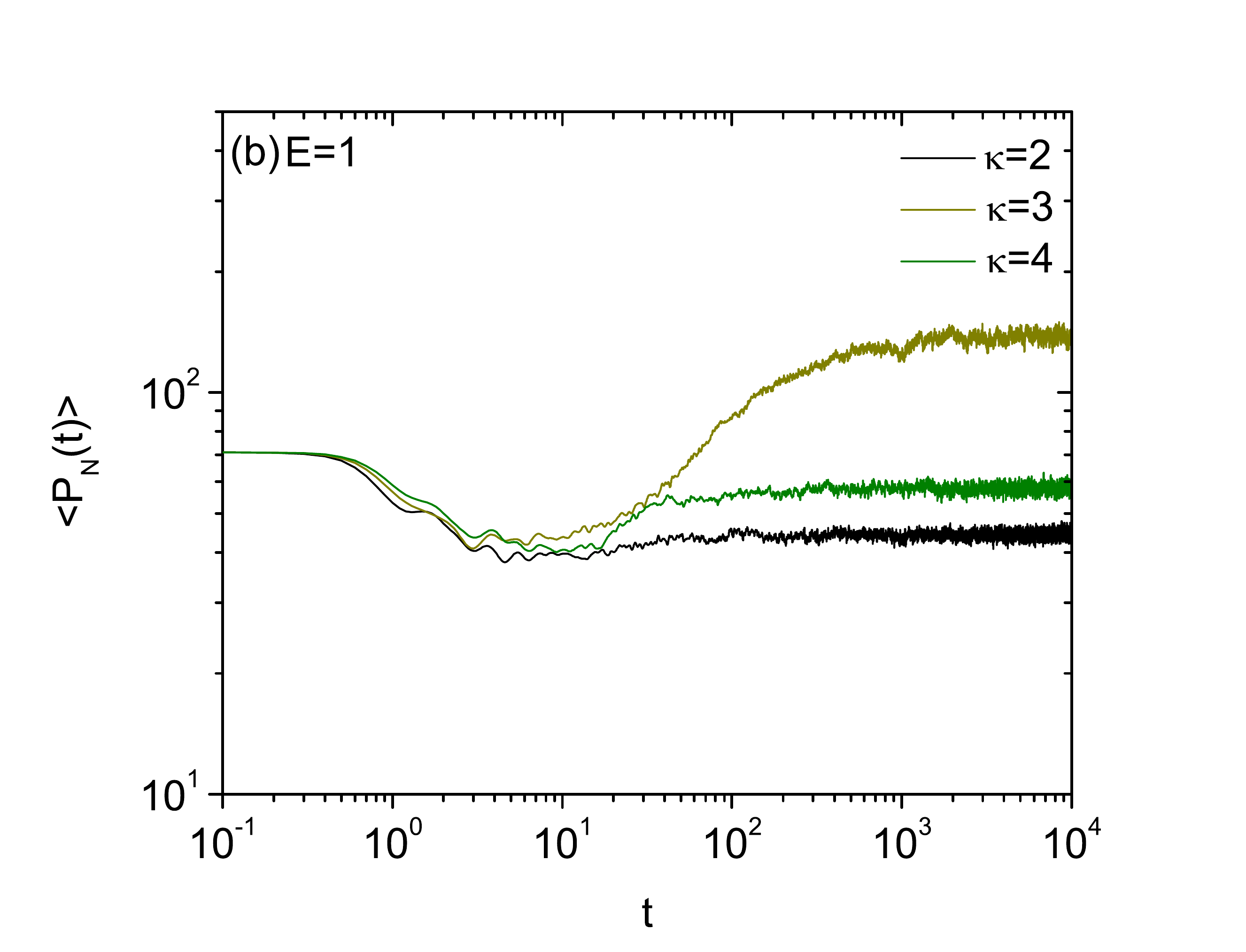}
\includegraphics[width=9cm]{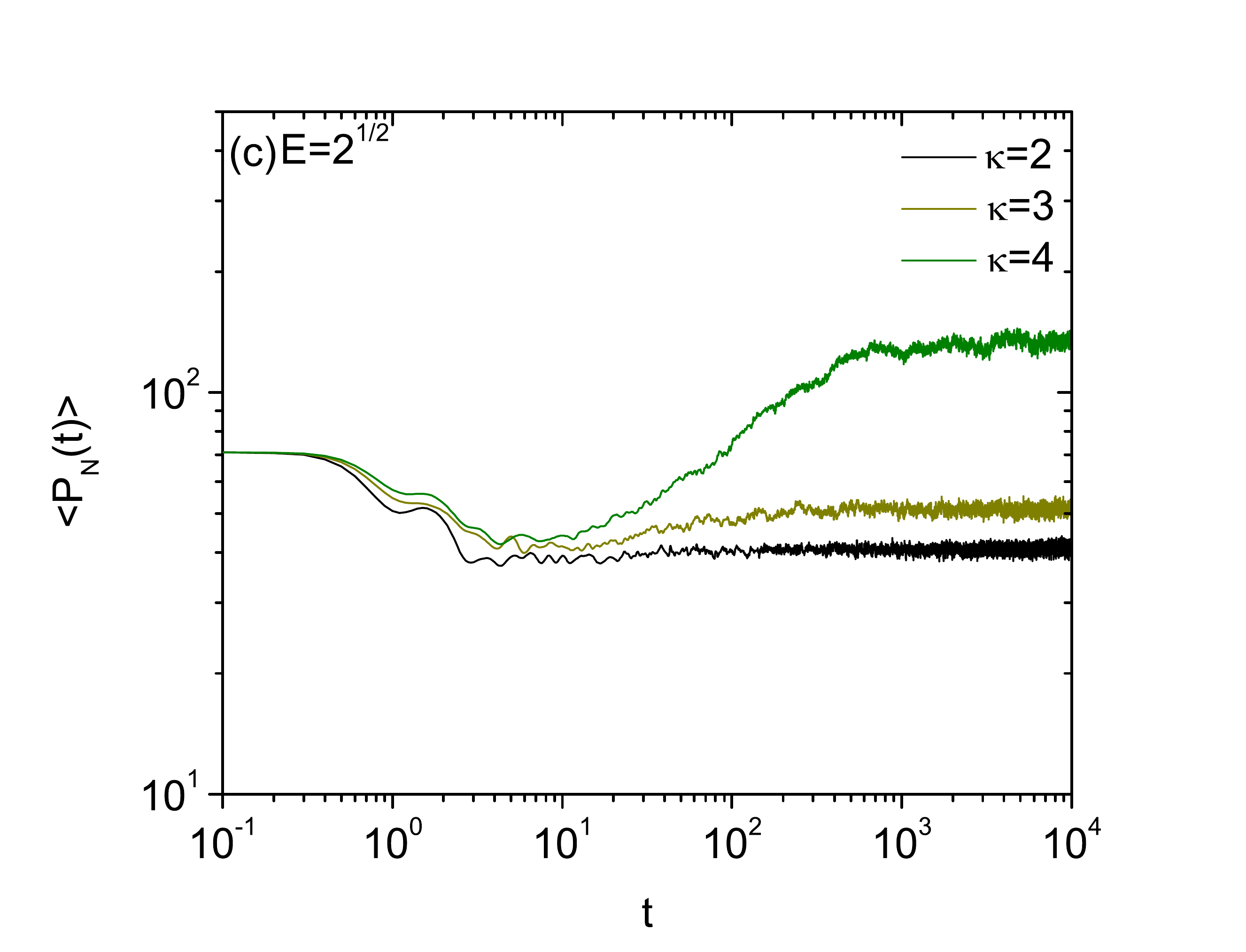}
\caption{Disorder-averaged participation number $\langle P_N(t) \rangle$ versus time $t$ when $\kappa=2$, 3, 4, $W=2$, $V_0=0$, and (a) $E=0$, (b) $E=1$, (c) $E=\sqrt{2}$. $\langle P_N(t) \rangle$ reaches an asymptotically finite value at the quasi-resonance energies at which $\langle m^2(t) \rangle$ increases as a power law in time.}
\label{fig8}
\end{figure}

\begin{figure}
\centering
\includegraphics[width=9cm]{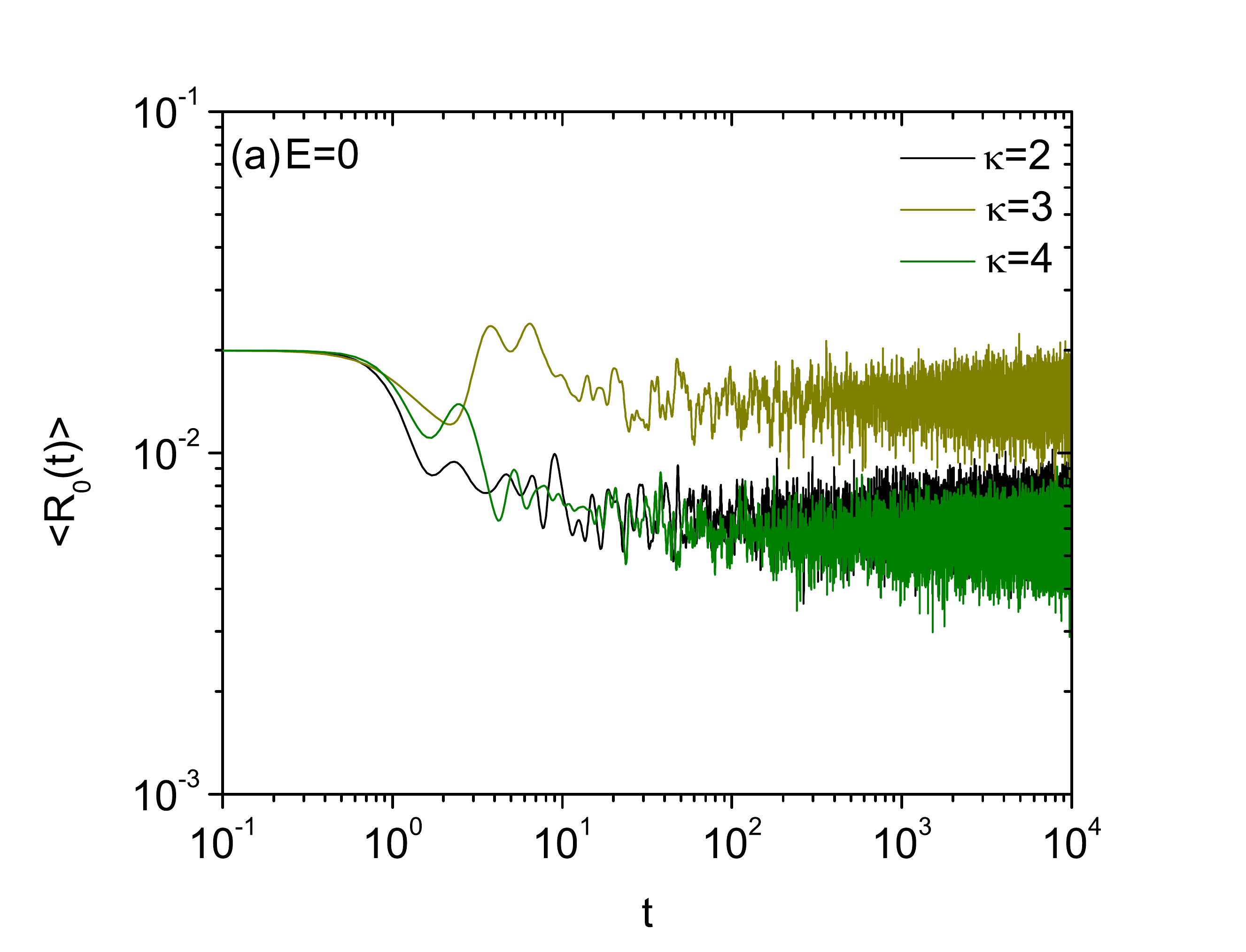}
\includegraphics[width=9cm]{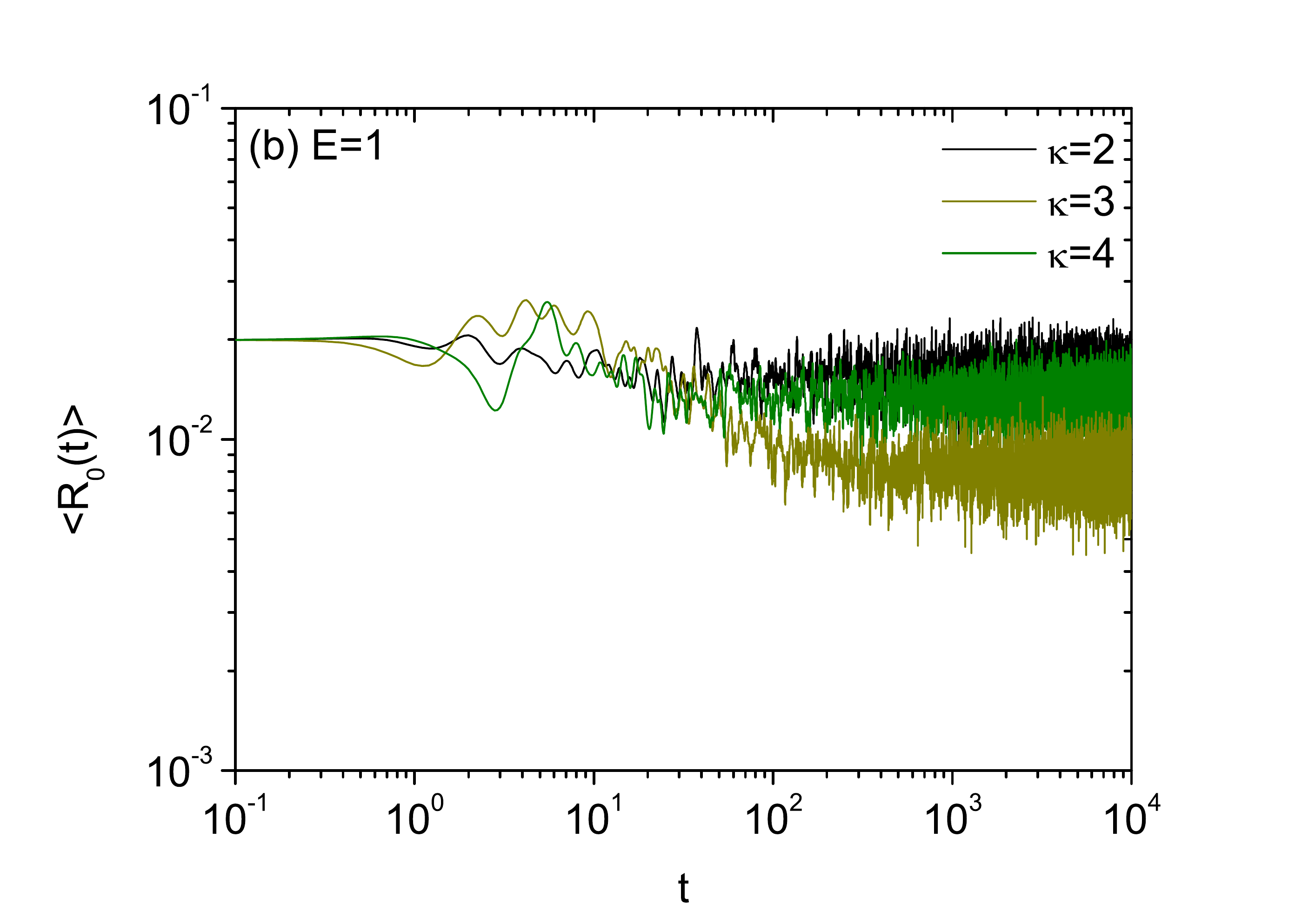}
\includegraphics[width=9cm]{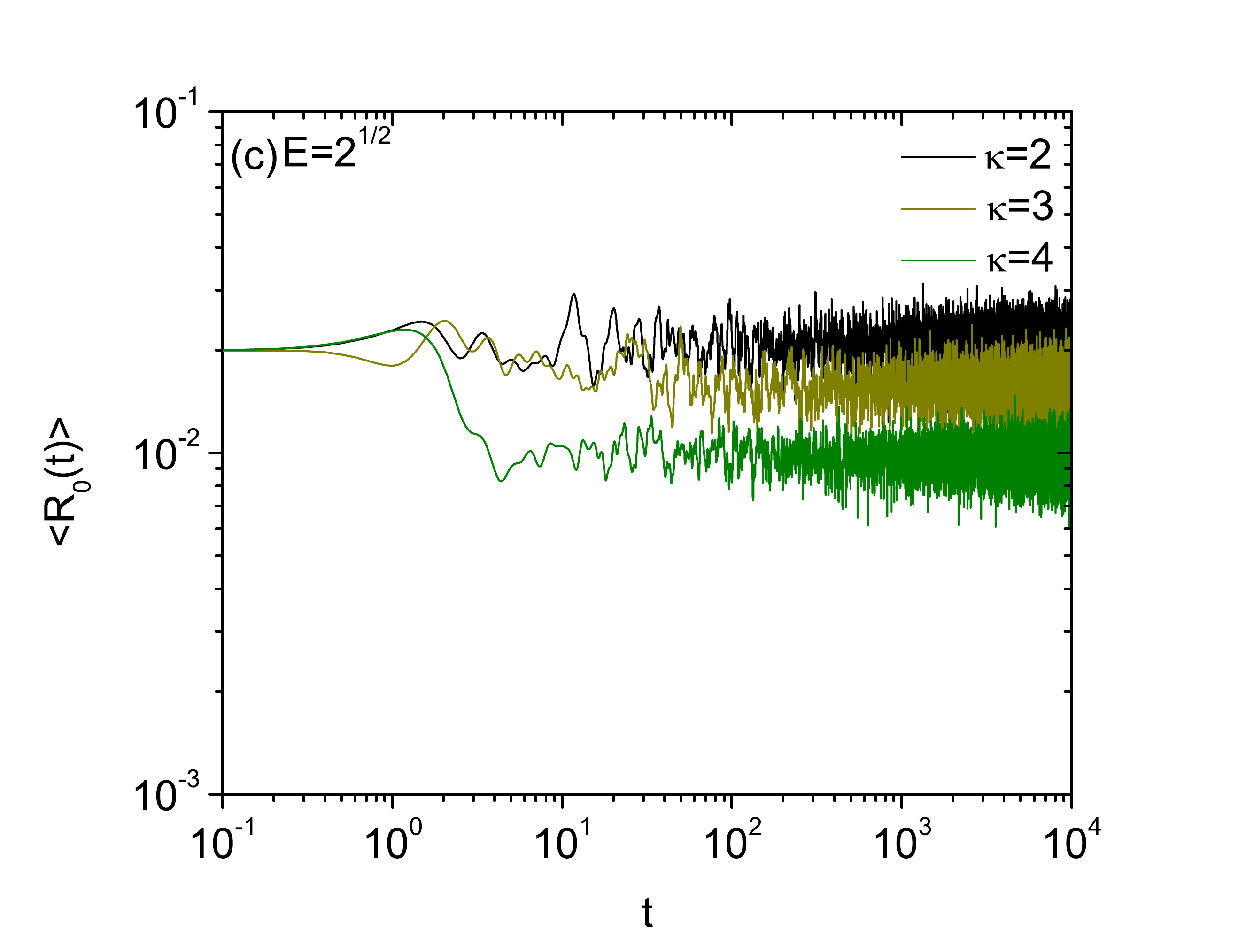}
\caption{Disorder-averaged return probability $\langle R_{0}(t) \rangle$ versus time $t$ when $\kappa=2$, 3, 4, $W=2$, $V_0=0$, and (a) $E=0$, (b) $E=1$, (c) $E=\sqrt{2}$. It is observed that there is always a finite probability of finding the quasiparticle at the initial position as $t\rightarrow \infty$.}
\label{fig9}
\end{figure}

We now consider the participation number which gives an estimation of the number of lattice sites where the wave packet has a significant amplitude \cite{Joh},
\begin{eqnarray}
P_N(t)=\frac{1}{\sum_{n=1}^{L}|C_{n}(t)|^4}
\label{equation16},
\end{eqnarray}
and the return probability defined by
\begin{eqnarray}
R_{0}(t)=|C_{n_{0}}(t)|^2.
\label{equation17}
\end{eqnarray}
A quasiparticle is said to have escaped from its initial position when the return probability $R_{0}(t)$ vanishes in the long-time limit. In contrast, the value of $R_{0}(t\rightarrow \infty)$ remains finite for a localized wave packet.
It has been known that the width of the spatial probability distribution of a wave packet is closely related to the mean-square displacement, while its amplitude is closely associated with the participation number and the return probability \cite{Lar}. In figures~\ref{fig8} and \ref{fig9}, we show the disorder-averaged participation number $\langle P_N(t) \rangle$ and return probability $\langle R_{0}(t) \rangle$ as a function of time $t$. All the numerical calculations are performed with the same parameters as in figure~\ref{fig6}. The results show that a partial localization phenomenon occurs at all the quasiresonance energies.
A signature of the presence of partial localization is
a saturation of $\langle P_N(t\rightarrow \infty) \rangle$ to a finite value as in the cases of $\kappa=2$ and 4 in figure~\ref{fig8}(a),
though the corresponding values of $\langle m^2(t) \rangle$ keep increasing owing to the expanding edges as $t\rightarrow \infty$ as shown in figure~\ref{fig6}(a). Similarly, asymptotically finite values of $\langle R_{0}(t\rightarrow \infty) \rangle$ are seen clearly in figure~\ref{fig9}. The observed behavior is not limited to energies where regular Anderson localization takes place but is also evident at quasiresonance energies, where subdiffusive transport is observed. All of these findings are in line with the existence of central peaks detected in the spatial probability distribution.

\section{Conclusion}
\label{sec41}

In this paper, we have expanded our previous research on the disordered mosaic lattice model
to perform detailed numerical calculations on various other physical quantities.
Our earlier study focused on exploring the delocalization effect that arises at a finite number of
quasiresonance energies primarily by analyzing the behavior of time-dependent reflectance of incoming wave packets over a long time interval. In this study, we have
examined the nature of states at quasiresonance energies through calculations on
the stationary quantities such as the disorder-averaged transmittance, logarithm of transmittance, and
participation ratio, $\langle T\rangle$, $\langle\ln T\rangle$, and $\langle P\rangle$,
and the dynamic quantities such as  the mean square displacement $\langle m^2(t) \rangle$, spatial probability distribution, participation number, and return probability.
For excitations at quasiresonance energies, we have found power-law scaling behaviors of the form $\langle T \rangle \propto L^{-\gamma_{a}}$, $\langle \ln T \rangle \approx -\gamma_g \ln L$, and $\langle P \rangle \propto L^{\beta}$.
Furthermore, when the wave packet's initial momentum satisfies the quasiresonance condition, we have
observed a subdiffusive spreading of the wave packet, characterized by $\langle m^2(t) \rangle\propto t^{\eta}$ where $\eta$ is always less than 1. It is known that critical modes of complex systems with multifractal spectra are associated with subdiffusive transport \cite{Geisel,Gua}. Exploring the potential connection between the multifractality of wave functions in real space and subdiffusive transport presents an intriguing avenue for future investigation.
We have also noted the occurrence of partial localization at quasiresonance energies, as indicated by the saturation of the participation number and a nonzero value for the return probability at long times.
The disordered mosaic lattice model studied in this paper can be readily realized experimentally using various physical systems, which include coupled optical waveguide arrays, synthetic photonic lattices, and ultracold atoms.
We hope that the results presented here will be a useful contribution to the study of unconventional localization phenomena in disordered systems.

\section*{Acknowledgments}
This research was supported through a National Research Foundation of Korea Grant (NRF-2022R1F1A1074463) funded by the Korean Government. It was also supported by the Basic Science Research Program funded by the Ministry of Education (2021R1A6A1A10044950) and by the Global Frontier Program (2014M3A6B3063708).

\end{document}